\documentclass{article}

\usepackage{microtype}
\usepackage{graphicx}
\usepackage{subfigure}
\usepackage{booktabs} % for professional tables
\usepackage{caption}
\usepackage{hyperref}

\usepackage[accepted]{icml2024}

\usepackage{amsmath}
\usepackage{amssymb}
\usepackage{mathtools}
\usepackage{amsthm}
\usepackage{multicol}
\usepackage{multirow}

\usepackage[capitalize,noabbrev]{cleveref}

\theoremstyle{plain}

\theoremstyle{definition}

\theoremstyle{remark}

\usepackage[disable,textsize=tiny]{todonotes}

\usepackage[normalem]{ulem}
\usepackage{xcolor}
\usepackage{listings}
\usepackage{subcaption}

\definecolor{commentgreen}{rgb}{0,0.6,0}
\definecolor{mylilac}{rgb}{0.58,0,0.82}

\lstdefinestyle{custombash}{
    language=bash,
    showstringspaces=false,
    basicstyle=\ttfamily,
    keywordstyle=\bfseries\color{green},
    commentstyle=\color{commentgreen},
    morecomment=[l][\color{commentgreen}]{\#}
}
\lstdefinestyle{customcpp}{
    language=c++,
    basicstyle=\ttfamily\footnotesize,
    commentstyle=\color{commentgreen},
    breakatwhitespace=false,
    breaklines=true, 
    showstringspaces=false,
    keywordstyle={\bfseries\color{blue}},
}
\lstdefinelanguage{wat}{
  keywords={module, type, param, local, result,  func, export},    % List of your keywords
  morekeywords={keyword3, keyword4, ...}, % Additional keywords
  sensitive=false,                        % Keywords are case-sensitive?
  morecomment=[l]{;;},                    % Line comment style
  morecomment=[s]{/*}{*/},                % Block comment style
  morestring=[b]",                        % String style
}
\lstdefinestyle{customwat}{
    language=wat,
    basicstyle=\ttfamily\footnotesize,
    commentstyle=\color{commentgreen},
    morecomment=[l][\color{commentgreen}], 
    breakatwhitespace=false,
    breaklines=true, 
    showstringspaces=false,
    keywordstyle={\bfseries\color{blue}},
}

\newcommand{\etal}{\textit{et al}. }

\icmltitlerunning{StackSight: Unveiling WebAssembly through LLMs and Neurosymbolic Chain-of-Thought Decompilation}

\begin{document}

\twocolumn[
\icmltitle{StackSight: Unveiling WebAssembly through Large Language Models and Neurosymbolic Chain-of-Thought Decompilation}



\icmlsetsymbol{equal}{*}

\begin{icmlauthorlist}
\icmlauthor{Weike Fang}{equal,usc}
\icmlauthor{Zhejian Zhou}{equal,usc}
\icmlauthor{Junzhou He}{usc}
\icmlauthor{Weihang Wang}{usc}
\end{icmlauthorlist}

\icmlaffiliation{usc}{Department of Computer Science, University of Southern California, Los Angeles, CA, United States}
%\icmlaffiliation{comp}{Company Name, Location, Country}
%\icmlaffiliation{sch}{School of ZZZ, Institute of WWW, Location, Country}

\icmlcorrespondingauthor{Weihang Wang}{weihangw@usc.edu}
%\icmlcorrespondingauthor{Firstname2 Lastname2}{first2.last2@www.uk}

\icmlkeywords{Machine Learning, ICML}

\vskip 0.3in
]



\printAffiliationsAndNotice{\icmlEqualContribution} % otherwise use the standard text.

\begin{abstract}
WebAssembly enables near-native execution in web applications and is increasingly adopted for tasks that demand high performance and robust security. However, its assembly-like syntax, implicit stack machine, and low-level data types make it extremely difficult for human developers to understand, spurring the need for effective WebAssembly reverse engineering techniques.
In this paper, we propose StackSight, a novel neurosymbolic approach that combines Large Language Models (LLMs) with advanced program analysis to decompile complex WebAssembly code into readable C++ snippets. StackSight visualizes and tracks virtual stack alterations via a static analysis algorithm and then applies chain-of-thought prompting to harness LLM's complex reasoning capabilities. 
Evaluation results show that StackSight significantly improves WebAssembly decompilation. Our user study also demonstrates that code snippets generated by StackSight have significantly higher win rates and enable a better grasp of code semantics.

% StackSight hinges on three crucial steps in WebAssembly decompilation: (1) explicitly tracking the virtual stack (2) recovering semantically meaningful variable names (3) articulating code functionality in natural language. 

% This dual approach aids in accurately predicting variable semantics, deciphering function objectives, and effectively decompiling WebAssembly code. 
% \textcolor{blue}{The necessity for such an advanced methodology also arises partly from the inherent scarcity of WebAssembly data in the datasets typically used to pretrain large language models.}

%\textcolor{blue}{This advancement not only showcases StackSight's effectiveness, but also underscores its contribution towards eliciting binary decompilation capabilities for low-resource assembly languages.} \ww{Could you elaborate what do you mean by ``low-resource" here?}
\end{abstract}

\section{Introduction}
WebAssembly (or WASM) is a low-level, portable bytecode language designed for high-performance computations at near-native speeds and broad portability across devices and platforms. It was first introduced for web applications~\cite{haas2017bringing} and is now supported by all major browsers, including Chrome, Firefox, Safari, and Edge~\cite{mcconnellnd}. It was then applied to a wide range of applications such as mobile devices~\cite{pop2022towards}, smart contracts~\cite{mccallum2019smartcontract}, and Internet of Things~\cite{gurdeep2019warduino,liu2021aerogel}.

Facilitating WebAssembly understanding is crucial, given its exploitation for malicious purposes such as cryptojacking, where it is executed secretly in browsers to mine cryptocurrencies~\cite{konoth2018minesweeper, kharraz2019outguard, musch2019thieves, romano2020minerray}. 
% However, WebAssembly is extremely hard to read for developers.
While WebAssembly serves as a compilation target for high-level languages such as C, C++, Go, and Rust, challenges arise when WebAssembly code is shipped as third-party modules without access to high-level source codes~\cite{musch2019new, romano2023automated}. 
Despite adopting a text format equivalent to WASM binary code, manual comprehension remains challenging. Different from register-based native binaries, WebAssembly manages a virtual stack machine, necessitating the tracking of stack behaviors to comprehend its operations. Furthermore, WebAssembly employs only four numeric data types (\texttt{i32}, \texttt{i64}, \texttt{f32}, and \texttt{f64}), concealing variable types and semantics. To make matters worse, 28.8\% of WebAssembly binaries are minified~\cite{hilbig2021empirical} with the variable names obfuscated and web test suites can get flaky~\cite{liu2024wefix}, making it hard to interpret and test. Such challenges motivate \textit{automatic decompilation} for better understanding.

% multi-step
Although \textit{decompilation} for WebAssembly could be naturally formulated as an end-to-end sequence translation task that takes assembly code as input and outputs the respective C++ source code, we find this naive formulation unsatisfactory in terms of performance and insight. The dissatisfaction arises from the relative scarcity of WebAssembly pretraining corpus and the fuzziness of neural networks.

In light of the trend in Natural Language Processing that encourages intermediate reasoning steps, we approach the decompilation problem in a step-by-step fashion.
We identify intermediate tasks humans would perform in order to decompile WebAssembly code. In correspondence with WebAssembly specifications, we propose three key stages: (1) explicitly tracking the virtual stack, (2) recovering semantically meaningful names of variables, and (3) summarizing code functionality in natural language. 
%These stages resonate with human logic and understanding. 
% We empirically demonstrate its effectiveness and constructiveness towards successful decompilation. 

In this work, we propose \textit{StackSight}, the first approach utilizing LLM with advanced program analysis for WebAssembly decompilation. 
StackSight breaks decompilation down into multiple smaller tasks: the program analysis component tracks the virtual stack as additional contextual information, and the chain-of-thought prompting ensures that the model can progressively interpret binary codes. This dual approach aids in accurately predicting variable semantics, deciphering function objectives, and effectively decompiling WebAssembly code. 
Besides, the necessity for such an advanced methodology arises partly from the scarcity of WebAssembly data in the datasets typically used to pretrain large language models, making them fail to grasp the nuances of this assembly-level language. We demonstrate that without StackSight, large language models perform poorly on WebAssembly comprehension and decompilation.

Existing literature has extensively demonstrated the capabilities of LLMs in various code reasoning tasks. However, our research uniquely directs its focus towards binary decompilation, which remains largely understudied within the community~\cite{fan2023large}. Beyond these primary tasks, our approach can be extended to downstream tasks such as malware detection and code review.
%StackSight-equipped models, on the other hand, almost doubles the percentage of correct decompilation.

% To validate our approach for real users, we conducted a user study focusing on the functionality and semantic similarity of the decompiled code. The first aspect evaluates how accurately and intuitively our method preserves the functionality of the original source code, while the second ensures the code resembles the semantics of the source code.

Specifically, our work makes the following contributions:
\begin{itemize}
    \item We develop a robust static analysis tool that explicitly tracks the state of the virtual stack, accounting for complex control and data flows. Equipped with the tool, we propose a novel Chain-of-Thought pipeline catered for WebAssembly decompilation.
    \item We perform comprehensive experiments on StackSight. Results show that StackSight increases the amount of functionally correct decompiled codes by 70\% and produces code that is notably more favorable from the perspective of human developers. 
    \item Case studies show that the static analysis tool mitigates logical errors and hallucinations in LLM outputs. 
\end{itemize}
\section{Related Work}

\subsection{WebAssembly Decompilation} 
In the realm of WebAssembly decompilation, significant efforts have been made towards recovering data and function types.
Lehmann and Pradel leveraged LSTM neural networks to recover precise, high-level parameter and return data types for WebAssembly functions~\yrcite{lehmann2022finding}. Romano and Wang generated semantics-aware intermediate representations of WebAssembly functions and applied machine learning classifiers to understand module and function purposes~\yrcite{romano2023automated}. These methods can significantly enhance human understanding of WebAssembly code, yet they fall short in unveiling the variety of code functionalities or enabling decompilation into high-level C/C++ codes. 

Benali was among the first to train a WebAssembly decompiler based on transformer and LSTM models~\yrcite{benali2022initial}. However, their method only works on artificial, small code snippets designed to evaluate different C grammars, lacking in real-world semantic complexity. Our approach, on the other hand, can work for more complicated WebAssembly code, identify diverse purposes, and decompile them to near-equivalent C/C++ code. This represents a significant leap forward in WebAssembly decompilation.

\subsection{Decompilation Using Static Analysis}
Researchers have also utilized traditional static analysis approaches to decompile binaries to C/C++ ~\cite{fokin2011smartdec, yakdan2015no, wang2017ramblr}, Java~\cite{desnos2011android,harrand2020java}, and Python~\cite{ahad2023pyfet}. A large body of work have focused on recovering Control Flow Graph (CFG)~\cite{Cifuentes1993ASA,cifuentes1995decompilation, fokin2011smartdec, yakdan2015no, wang2015reassemble,wang2017ramblr, gussoni2020comb} and type inference from binary code~\cite{lee2011tie,elwazeer2013scalable,noonan2016polymorphic,xu2017learning,lehmann2022finding}.  Liu and Wang performed a comprehensive study to investigate decompilation
correctness of C code decompilers, finding that while modern decompilers have been progressively improved, challenges such as type recovery and
optimization still hinder well-formed outputs ~\yrcite{liu2020far}. 
Our approach distinctively integrates static analysis with a focus on the virtual stack machine to infer variable types and relationships, thus tackling the inherent complexities of WebAssembly. 
Unlike traditional static analysis methods that may not fully consider WebAssembly's implicit memory mechanisms, our approach provides a detailed understanding of these elements to reconstruct high-level data types and semantics from the binary code.

\subsection{Neural Decompilation}
Prior work has leveraged Neural Machine Translation (NMT) models for binary decompilation~\cite{katz2019towards, liang2021semantics}. Katz \etal presented a Recurrent Neural Network (RNN) based method to decompile machine code to high-level C code~\cite{katz2018using, fu2019coda}.
Cao \etal targeted compiler-optimized binaries and leveraged graph neural network (GNN) model to convert binaries to an intermediate representation (IR), which is then used to generate high-level code~\yrcite{cao2022boosting}. 
% Szafraniec \etal leverage LLVM Intermediate Representations (IR) to improve neural machine translation for source code~\cite{szafraniec2023code}. 

Recently, Large Language Models have shown impressive performance on code-related tasks, particularly in binary reverse engineering~\cite{pearce2022pop,xu2023lmpa, wong2023refining,alkaswan2023summarize}. 
% Prior works using LLMs for binary reverse engineering have demonstrated promising results~\cite{xu2023lmpa, pearce2022pop,wong2023refining,alkaswan2023summarize}. 
All these efforts aim to decompile C/C++ executable binaries back to C/C++. Our work stands at the intersection of these neural decompilation methods and the unique challenges presented by WebAssembly binaries. This focus on WebAssembly sets our work apart, as we address the specific nuances and complexities inherent in these binaries, contributing novel insights and methodologies to the field of decompilation.
%All the efforts are to decompile C/C++ executable binaries back to C/C++. We study decompiling WebAssembly binaries, which are emerging prevalent in web applications, back into C/C++ codes with the use of LLM.

%\textcolor{red}{Add more LLM related works.}

\section{Methodology}
\begin{figure*}[h]
    \centering
    % \vspace{-0.5em}
    \includegraphics[width=1.9\columnwidth]{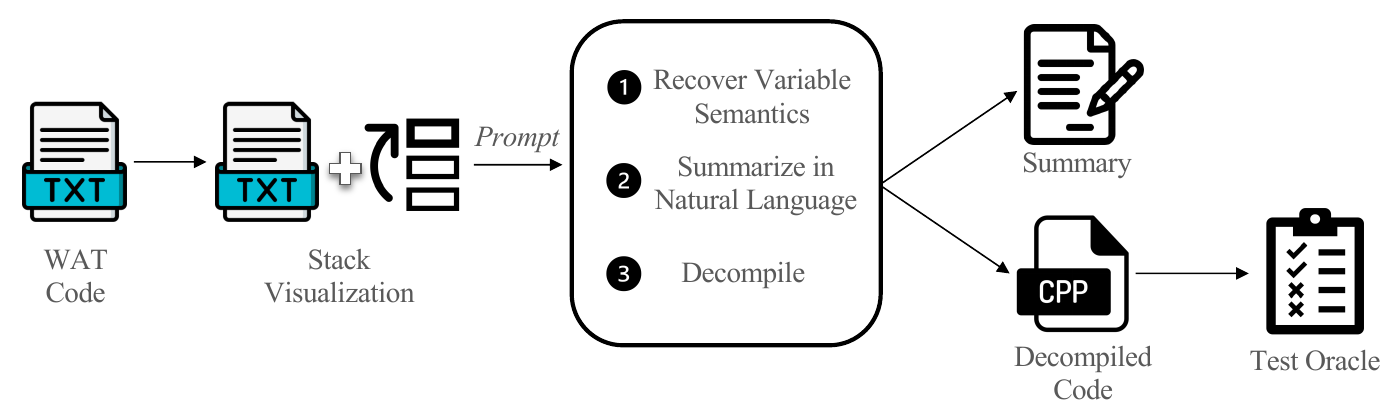}
    \caption{StackSight: The decompilation pipeline.}
    \label{fig:pipeline}
    %\vspace{-1em}
\end{figure*}
\subsection{Motivations behind Step-By-Step Decompilation}

In our methodology, we draw inspiration from recent advancements in eliciting reasoning abilities in LLMs. One major advancement in the NLP community is the Chain-of-Thought (CoT) Prompting approach, first introduced by~\cite{wei2022chain}. CoT prompting focuses on guiding large language models through a series of intermediate reasoning steps resembling the thinking process of a human, significantly improving the ability of LLMs to perform complicated reasoning. 

Our work is the first to propose using CoT reasoning in reverse engineering tasks like decompilation. Previous work~\cite{benali2022initial} has shown that a direct sequence-to-sequence model from WebAssembly binaries to high-level C code fails to grasp such complex data flow and generate high-quality source code when code complexity increases. 
It is intuitive to decompose the decompilation process into multiple phases as a large amount of complex reasoning is needed to interpret, abstract, and translate binary code to human-readable languages. We reflect on how humans would approach the WebAssembly decompilation task and strategically decompose the process into multiple steps, eliciting LLMs' reasoning abilities to decompile like a human.
We identify the following three intuitions that are critical to unveil the assembly-like nature of WebAssembly code.

\noindent \textbf{Explicitly Tracking the Virtual Stack}\quad As specified in the original paper~\cite{webassembly}, WebAssembly code operates on an implicit stack. An example WebAssembly Text Format file is provided in Listing \ref{lst:wat}.

As shown, WebAssembly code loads variables onto the stack (\lstinline[style=customwat]{local.get, i32.const}), operates on the top stack elements and writes back the answer (\lstinline[style=customwat]{i32.add, i32.mul, i32.div_s}). However, the text format lacks explicit representation tracking the history states of the stack.

While this approach can achieve a more compact code size, it also imposes great difficulties for humans to understand the code. We concur that keeping an explicit history of what is on the virtual stack and the operations that interact with the stack would serve as a useful intermediate step. More than helping to identify the semantics of WebAssembly code, this step also reveals patterns in \texttt{.wat} files that correspond to specific behavior in C++ source code. 

\begin{lstlisting}[style=customwat, label={lst:wat}, caption={Example of WebAssembly Text file (WAT).},captionpos=b]
(func (;1;) (type 1) (param i32) (result i32)
    local.get 0
    i32.const 1
    i32.add
    local.get 0
    i32.mul
    i32.const 2
    i32.div_s)
\end{lstlisting}
% (module
%(type (;0;) (func))
%(type (;1;) (func (param i32) (result i32)))
%(func (;0;) (type 0))
% % (export "__wasm_call_ctors" (func 0))
% (export "__wasm_apply_data_relocs" (func 0))
% (export "_Z8sum_to_ni" (func 1)))

\noindent \textbf{Recovering Semantically Meaningful Names of Variables}\quad Also specified in the original paper~\cite{webassembly}, WebAssembly code only assigns indexes to used variables, posing another major obstacle to understanding WebAssembly code. Human developers typically encode semantic information in variable names. To recover variable semantics in WebAssembly code, we first prompt the LLM to predict the variable types of input and return values based on stack information. Then, the model gradually identifies the purpose of each local variable by inspecting operation histories and assigns meaningful names to each variable.
%Therefore, we also agree on recovering semantically meaningful names for used variables. 
%This intuition is also supported by our discovery that the WebAssembly compiler follows certain conventions when assigning local variables. One type of information tends to be stored at a specific index.

\noindent \textbf{Summarize in Natural Language}\quad Once all variable semantics are recovered, the next logical step for human developers to interpret the WebAssembly code is to identify high-level purposes of the code. 
This is crucial in preventing the verbatim transcription of WebAssembly instructions into potentially incorrect C/C++ code. Our approach focuses on a natural language interpretation of each part of the code, which provides a more intuitive understanding of the code's functionality, aligning with how developers typically conceptualize complex coding structures.

\subsection{StackSight: WebAssembly Decompilation Pipeline}

To facilitate stepwise reasoning, we build a pipeline shown in Figure~\ref{fig:pipeline} for WebAssembly decompilation, where each component focuses on one of the above decompilation phases. This structured pipeline ensures that LLM can progressively build upon its understanding.
% \subsubsection{Translation to Text Format}
% Given a piece of binary WebAssembly code (\texttt{.wasm}), we first translate it to the WebAssembly text format (\texttt{.wat}) using \texttt{wasm2wat} tool from the WebAssembly Binary Toolkit (WABT)~\footnote{https://github.com/WebAssembly/wabt}.

\subsubsection{Stack Visualization}
In StackSight, we address the inherent complexities of WebAssembly's virtual stack machine through a specialized static analysis tool. This tool is adept at parsing the WebAssembly text format (WAT) code, thereby explicitly visualizing and tracking the behavior of the stack machine. 

WebAssembly diverges from traditional native binaries in that it permits the definition of any number of local variables and effectively operates with an unlimited number of virtual registers on a virtual stack machine. This dependency obfuscates the understanding of the code's functionality, both for humans and computational models, without prior knowledge of the stack's configuration.

To overcome this challenge, StackSight includes a static analysis algorithm to visualize the stack after each operation. This approach mirrors how a human reverse engineer may jot down the stack and interpret operations. The algorithm supports all 172 WebAssembly opcodes, including memory operations, function calls, type conversions, control flow, and branching operations. As each opcode has its specific rules on how it manipulates the stack, StackSight simulates such manipulations to precisely determine the stack's resulting state from any given instruction and its current state. Moreover, it applies symbolic execution to account for control flow branches: When there are divergent execution paths that may result in different stack states, it will use symbolic notation to record different branches.

\begin{figure*}[h]
    \centering
    % \vspace{-0.5em}
    \includegraphics[width=2\columnwidth]{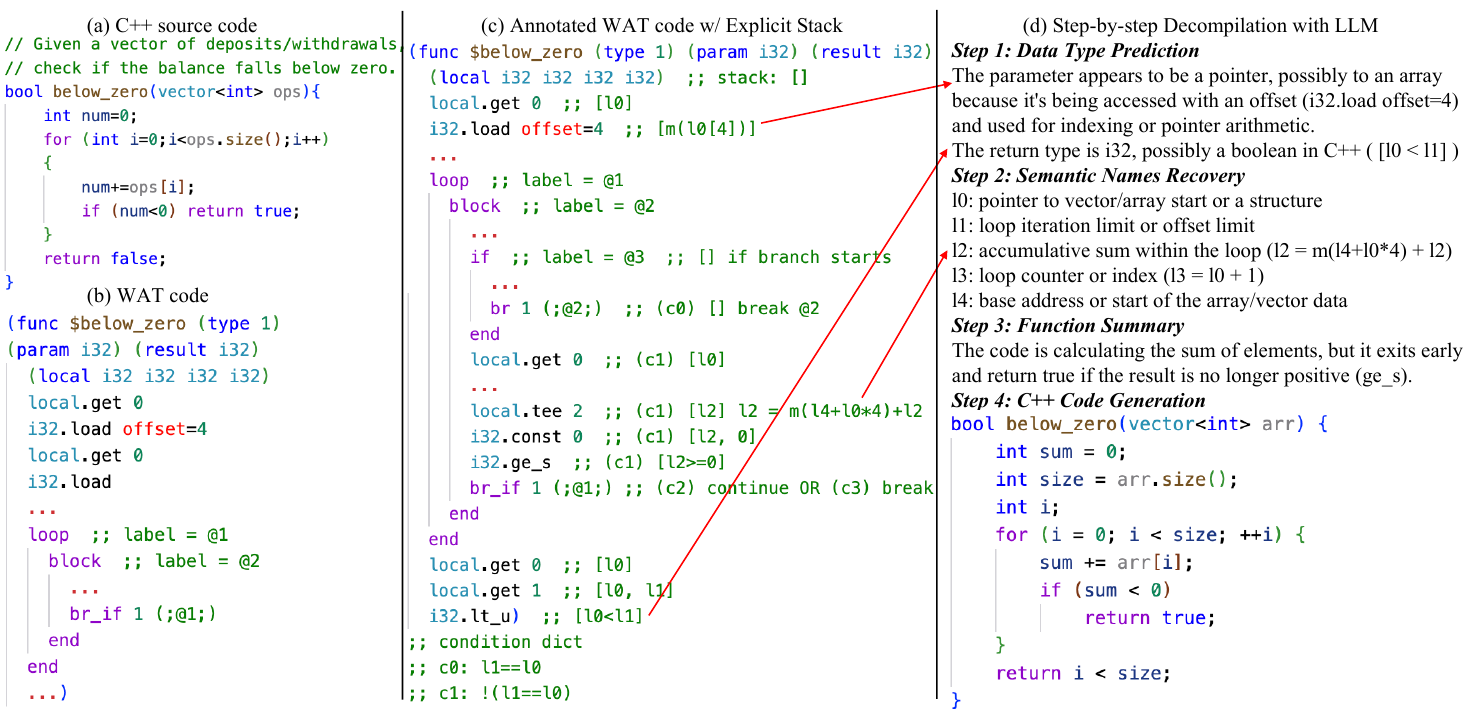}
    \caption{A Running example for StackSight. (a) C++ source code in HumanEval-X dataset. (b) WebAssembly Text Format (WAT) code compiled with Emscripten and translated with WABT. (c) Annotated WAT code with our proposed static analysis tool capturing the virtual stack alterations, variable assignments, control flow, and branching statements. (d) Chain-of-Thought prompting by breaking decompilation into multiple phases. LLM is able to utilize stack annotations, make reasonable predictions, summarize function purposes, and decompile WAT code to high-level C++. }
%\ww{The caption description is great! However, is it possible to simplify the code further? Also, try to align the subfigure captions (a), (c), and (d) with the same height.}}
    \label{fig:example}
    %\vspace{-1em}
\end{figure*}

Our static analysis tool can annotate each line of WAT code with the precise state of the stack. 
The exact state of the stack at each line of code is represented with a simplified notation.  This approach mirrors the steps a human reverse engineer might take, such as manually noting the stack's state to interpret its operations. An example can be seen in Figure~\ref{fig:example}. We show empirically that this program analysis tool can significantly bridge the gap between WASM code and its high-level source code counterparts.

\subsubsection{CoT Prompting}
The annotated WebAssembly code (\texttt{.wat}) is then included in the prompt to a large language model in a step-by-step manner. First, based on the stack visualization, we ask the LLM to comprehend variable relationships, predict the data types for input parameters and return types (\textit{pivot variables}). The model is then prompted to recover the semantic meaning of each local variable based on the pivot variables. Given the context and usage of each local variable, the model is asked to summarize the function operations in natural language to derive a high-level understanding of what the WebAssembly code is doing. Finally, based on the above knowledge, the LLM is prompted to decompile the piece of WebAssembly code to C or C++. To evaluate the outputs from the LLM, we implement a test oracle that incorporates a series of test cases to evaluate the correctness of the decompiled code. 
%Compilation errors and failed test cases are passed as feedback to the prompt generator to prompt the model for another round of generation. Such iterative process will improve the generation over time, and will terminate if the pass rate goes above a certain threshold or the number of iterations exceed the query budget. \textcolor{red}{Due to the nondeterministic nature of LLM such as GPT, we would ideally repeat each prompt for a number of times to get a more diverse set of outputs. Our robustness study in Section \ref{sec:robustness} shows that predictions are rather stable given the same query and seed. In practice, we employ the threshold of pass rate to be xx.}

% \input{tex/intuition}
% \input{tex/baselines}
\section{Experiment Setup}
\subsection{Datasets}
\begin{table*}[t]
\centering
\caption{Data statistics for HumanEval-X and MBXP benchmarks from \cite{nl2code}, with modifications. \textit{Num.} denotes the number of instances in the benchmark, \textit{Working Num.} denotes the number of instances that can compile,  \textit{S.PL} denotes code Solution's Programming Language (we are using the C++ split),  \textit{T.N.} denotes the average Number of Test cases per function,  \textit{W.C.} and \textit{W.L.} for the average number of Characters and Lines in compiled WAT file,  \textit{S.C.} and \textit{S.L.} for the average number of Characters and Lines in canonical Solution.
% \ww{The caption of other tables stays above the table, while this one stays below; please double-check the template and make sure the format is correct.}
}
% \footnotesize
\setlength\tabcolsep{5pt}
\renewcommand{\arraystretch}{1.1}
\begin{tabular}{lcccccccccc}
\toprule
Benchmark & Num. & Working Num. & S. PL & T/N. & W/C. & W/L. & S/C. & S/L. & Scenario \\
\midrule
HumanEval-X (2023) & 164 & 161 &  C++ & 7.8 & 6784.9 & 295.2 & 252.5 & 10.4 & Code Exercise \\
MBXP (2022)        & 974 & 773 &  C++ & 3.1 & 6063.9 & 160.1 & 192.9 & 9.2    & Code Exercise \\
\bottomrule
\end{tabular}
\label{tab:datasets}
\end{table*}
% \begin{table*}[t]
% \centering
% \begin{tabular}{lcccccccccc}
% \toprule
% Benchmark & Num. & Working Num. & P. NL & S. PL & T/N. & P/C. & P/L. & S/C. & S/L. & Scenario \\
% \midrule
% HumanEval-X (2023) & 164 & 161 & English & C++ & 7.8 & 468.4 & 15.5 & 264.6 & 12.1 & Code Exercise \\
% MBXP (2022)        & 974 & 761 & English & C++ & 3.1 & 419.9 & 14.8 & -     & -    & Code Exercise \\
% \bottomrule
% \end{tabular}
% \caption{Data statistics for HumanEval-X and MBXP benchmarks from \cite{nl2code}, with modifications. \textit{Num.} denotes the number of instances in the benchmark, \textit{Working Num.} denotes the number of instances that can compile, \textit{P.NL} denotes Problem description's Natural Language, \textit{S.PL} denotes code Solution's Programming Language (we are using the C++ split), and \textit{T.N.} denotes the average Number of Test cases per function. \textit{S.C.} and \textit{S.L.} (\textit{P.C.} and \textit{P.L.}) stand for the average number of Characters and Lines in Code Solution (Problem description).}
% \label{tab:datasets}
% \end{table*}

We list our main considerations when choosing evaluation datasets. First, we need to ensure that the datasets are compilable to WebAssembly, to obtain input-output pairs. Second, we need to ensure that the datasets come with test cases and natural language descriptions to assess \textbf{functional correctness} of decompiled code snippets and the \textbf{semantic similarity} of model-generated summaries. Finally, we minimize data contamination concerns by verifying that the datasets are human-written. 
To this end, we use the two code benchmark datasets. Both datasets are originally targeted for Natural Language to Code generation. Each sample comes with test cases and a natural language task description that satisfies our needs. For examples of the two datasets, see Appendix~\ref{app:dataset_examples}

\noindent \textbf{HumanEval-X}\quad HumanEval-X~\cite{humaneval_x} is a widely-used Natural Language to Code dataset which has a split in C++. It is adapted from   HumanEval~\cite{humaneval} by human experts. The HumanEval dataset is also written by human experts, therefore reducing the risk of data contamination. 
%For an example of the HumanEval-X dataset, see Appendix~\ref{app:humanevalx_example}.

% Under the original setting of HumanEval-X, the model is prompted to complete a single function based on the 

\noindent \textbf{MBXP}\quad MBXP~\cite{mbxp_dataset} is also a Natural Language to Code dataset that contains a C++ split. It is created using a parsing-based conversion framework that generates C++ code from Python snippets in MBPP~\cite{mbpp_dataset}. The MBPP dataset is crowdsourced, therefore also deemed to be original. 
%For an example of the MBXP dataset, see Appendix~\ref{app:mbxp_example}.
% It focuses on \textit{entry-level} tasks for programmers.
\subsection{Compilation Pipeline}
We use Emscripten~\cite{emscripten} (version 3.1.46) to compile C++ code to WebAssembly. We leverage the \texttt{wasm2wat} (version 1.0.33) tool from the WebAssembly Binary Toolkit repository~\cite{wabt} to convert binary instructions (\texttt{.wasm}) into text format (\texttt{.wat}).
% We keep the LLVM IR~\cite{lattner2004llvm} as a useful intermediate-level representation. 
To improve the readability of \texttt{.wat} files, we 1) compile function snippets as side modules 2) perform compiler optimization to reduce the size of generated code 3) preserve function names in C++ 4) present symbols in a readable way. The exact commands that we use are in Appendix \ref{compilation_cmd}. For a summary on our compiled datasets, please refer to Table~\ref{tab:datasets}.
% emcc "$target" -o "$output_ll" \
%     -S \
%     -emit-llvm \
%     -s SIDE_MODULE=1 \ # compile as side module
%     -Oz # optimize size

% \subsection{Summary on Compiled Datasets}
% After filtering out examples that 1) do not have canonical solutions, or 2) cannot be easily fixed to compile, we obtain \textbf{161} working examples for \textit{HumanEval-X} and \textbf{761} working examples for \textit{MBXP}.

\subsection{Models and Baselines}

\subsubsection{Baselines}
A baseline approach is to elicit LLM's In-Context Learning (ICL) ability ~\cite{dai2023gpt} by providing it with one or more examples as context. It is also known as one-shot or few-shot learning, which involves presenting LLMs with one or a few examples to provide context and understanding of a specific task. Such a method is widely applied for its simplicity and ability to quickly adapt to new tasks without the need for training or fine-tuning. 

We construct in-context learning examples by concatenating three parts. We first start with an instruction part to explicitly state that the task to perform is to decompile WebAssembly files into the corresponding C++ file. We then append one or several round(s) of user and AI system interaction to serve as an in-context learning example. The user input is an exemplar \texttt{.wat} file, while the system output is the corresponding C++ file it is compiled from. Finally, we append the target \texttt{.wat} file of interest. 
%\ww{Just curious, is this second .wat file different from the first one that will be decompiled?}
For an example prompt, please refer to Appendix~\ref{appendix:prompt_eg}. Other WebAssembly decompilers, such as \texttt{wasm2c} by WABT and \cite{benali2022initial}, fail to generate compilable codes for the two datasets and are thus not included in the evaluation.

As the static analysis step involves constantly tracking the virtual stack machine and emulating branching operations, we find that the annotated code is nearly twice as large as the original code in average. 
For a fairer comparison, we include more shots in the ICL baselines to match the prompt size. We include one shot in StackSight. As the annotated code is twice as large, 2-shot or 3-shot ICL baseline will match the prompt size of StackSight. To compare our method against stronger baselines, we randomly select examples from both datasets, and conduct the same set of experiments for 3-shot, 5-shot, and 10-shot ICL in our evaluation.

\subsubsection{Models}
We conduct experiments on three large language models: gpt-3.5-turbo-1106~\cite{openai2024gpt35turbo}, gpt-4-0125-preview~\cite{openai2024gpt4},  and Code Llama-7b-Instruct~\cite{rozière2023code}.
All three models have exhibited strong code interpretation and complex reasoning abilities. As WebAssembly is low-resource, and its decompilation is especially intricate, choosing less advanced models might compromise the quality and accuracy of the decompilation process.
Besides, their easy accessibility ensures that our research can be replicated and applied by a broader community of developers.

\section{Evaluation}

We examine the effectiveness of our pipeline by addressing the following research questions:

\textbf{RQ1:} To what extent do the decompiled codes generated by our method match their original codes?
% \textbf{RQ1:}  How well do decompiled code by our method and SOTA match their original versions in the source code?

\textbf{RQ2:}  Can our method more effectively help developers comprehend WebAssembly code in the real world?
% , and how does it compare to SOTA WebAssembly decompilers? 
% \ww{Did we test SOTA WebAssembly decompilers?}

\textbf{RQ3:}  How does each component of our pipeline contribute to the overall performance of decompilation?

% \textbf{RQ4:} Is it resilient to the nondeterminism of LLM answers?

% \textbf{RQ4:}  How well is our pipeline on unseen programs and real-world large project?

\subsection{RQ1: Quality of Decompiled Code and Summary} 
%However, our experiments reveal that In-Context Learning faces challenges in the reverse engineering tasks such as WebAssembly decompilation. Without the intermediate steps, it is hard for LLM to logically bridge the low-level WebAssembly code and the corresponding expected C/C++ code. This finding underscores the complexity of such decompilation tasks and the need for more specialized approaches. 

%The comparison serves to demonstrate the superiority of our approach in breaking the task down to a series of subtasks, thus accurately translating and preserving the functionality of WebAssembly code in its high-level C/C++ code representation.

In addressing this research question, our evaluation is twofold: assessing the quality of the generated summaries and examining the correctness of the decompiled code.

\noindent \textbf{Evaluating Summary Quality}\quad We measure the quality of the summaries produced by our pipeline using both automated text quality metrics and human judgment.

\noindent \textbf{Assessing Decompiled Code Correctness}\quad To evaluate the quality of the decompiled code, we choose \textbf{functional correctness} as our metric. Previous works on decompilation~\cite{benali2022initial, fu2019coda} have focused on metrics such as token-level accuracy, template-level accuracy and also complete match. Such metrics capture textual and lexical similarities.~\cite{humaneval} suggests that text-similarity based metrics do not correlate with functional correctness, and proposes to directly evaluate correctness through expert-written test cases. We follow this setup to quantify the quality of decompiled code.
We employ a test oracle to incorporate C/C++ test cases into the decompiled code and report the pass rate to assess the effectiveness of our decompilation pipeline.

% Our evaluation compares these aspects of our pipeline against the performance of state-of-the-art (SOTA) WebAssembly decompilers and In-context Learning (ICL) method.

\subsubsection{Summary Evaluation}
Evaluating the degree of alignment between the predicted summaries and ground-truth summaries provided in the dataset presents a challenge because they may be written in different styles or formats but are semantically similar to each other~\cite{fang2022investigating}. In order to capture semantic similarity, we propose BERTScore~\cite{zhang2020bertscore} and Sentence-BERT (SBERT)~\cite{reimers2019sentencebert} for evaluation. BERTScore leverages the contextual embeddings from BERT to compute the similarity between predicted and reference sentences. SBERT is designed for creating dense vector representation of sentences. Similarity scores are then computed using cosine similarity.

\subsubsection{Decompilation Correctness Evaluation}

% \noindent \textbf{Objective}\quad
% The primary goal of our decompilation pipeline is to convert WebAssembly code into functionally equivalent C++ code. Functional equivalence is defined as preserving core functionality but not necessarily syntactic elements such as function names and parameters. 

% \noindent \textbf{Approach}\quad
% To evaluate the functional fidelity of the decompiled C++ code, we design a test oracle that is able to insert test cases to the decompiled code and measure the pass rate to validate the functionality of decompiled code. 

% \noindent \textbf{Evaluation Steps}\quad
Using automatic parsers to extract desired model output and automate the evaluation process is a common practice in evaluating LLMs, as showcased in MetaMathQA~\cite{yu2023metamath} and MMIQC~\cite{liu2024augmenting}. We implement the test oracle to automate the evaluation, including five steps:
% \noindent \textbf{Summarize in Natural Language}\quad

\noindent \textbf{Code Sanitization}\quad The test oracle parses the outputs from LLMs and extracting the functions under test (FUT). In some cases, the generated code may include an extraneous \texttt{main()} function intended to validate the usage of the FUT. In such instances, the test oracle removes it, since test cases will be integrated into a new main function automatically. 

\noindent \textbf{Function Name Alignment}\quad Function names from decompiled code may divert from those expected by the test cases because WebAssembly compilers such as Emscripten may strip away function identifiers. The test oracle will rename the function names from the decompiled code to match those expected by the test cases, utilizing regular expression
%Therefore, we perform a renaming process to ensure that the function names in the decompiled code match exactly with those in the test oracle. 
%\ww{In section 4.2 compilation pipeline, "we compile functions and preserve function names in C++". Do we still need to align function names here?}

\noindent \textbf{Test Case Integration}\quad Subsequently, the test oracle will insert test cases pertinent to this function into the decompiled code in a new \texttt{main()} function. This step assesses whether the decompiled code meets the functional requirements as specified by the test suite. 

\noindent \textbf{Compilation Error Rectification}\quad We address compilation errors that do not impact the underlying functionality.

\textit{Function argument pass-by-reference}: Generated code from LLM often uses pass-by-reference for function arguments to minimize copying overhead. However, in the tests within both evaluation datasets, constant vectors are often passed to function arguments. To address this issue, we manually modify the function argument to use pass-by-value or prefix the \texttt{const} modifier when it does not alter the function's intended behavior. 

\textit{Namespace issues}: In some instances, the tests presuppose the inclusion of \texttt{using namespace std;} directive. This line is automatically inserted where necessary

%The most common issues encountered are: function argument mismatch, inclusion of libraries, and namespace issues. We fix these errors where necessary to ensure successful compilation.

\noindent \textbf{Test Case Execution}\quad Finally, we compile and execute the test cases against the decompiled code. The test oracle will output the percentage of decompiled files that can compile and those that can pass all test cases.

% \begin{table*}[htbp]
% \centering
% \caption{Summary evaluation and comparison of pass rate of different methods. For summary evaluation, we report BERTScore and SBERT. For decompilation quality, we report the percentage of outputs that can compile successfully and pass all test assertions.}
% \label{tab:correctness_comparison}
% \begin{tabular}{lcccccccc}
% \toprule
% & \multicolumn{4}{c}{HumanEval-X} & \multicolumn{4}{c}{MBXP} \\ 
% \cmidrule(lr){2-5} \cmidrule(lr){6-9}
% Metric & \% Compile & \% Correct & BERTScore & SBERT & \% Compile & \% Correct & BERTScore & SBERT \\ 
% \midrule
% GPT-3.5  & 38.2 & 13.4 & 80.89   & 38.27 & 48.0& 19.6& 83.31& 36.29\\
% + StackSight  & 58.1 & 31.4 & 83.64 & 51.82 & 54.9& 29.5& 84.58& 51.32 \\
% $\Delta$ & +19.9 & +18.1   & +2.74 & +13.55 & +6.9& +9.9& +1.27 & +15.03 \\
% \midrule
% Code Llama  & 47.7 & 13.5 &   79.48 & 37.10& 38.8& 10.3& 81.47& 37.00 \\
% + StackSight & 54.2 & 24.6  & 83.13 & 53.46 & 41.8& 15.4& 85.41& 55.04\\
% $\Delta$ & +6.5 & +11.1   & +3.66 & +16.36 & +3.0& +5.1& +3.94 & +18.04 \\
% \midrule
% GPT-4      & 46.5 & 19.7 & 78.75   & 33.85 & 63.6 & 42.4& 80.73& 34.50\\
% + StackSight  & 60.6 & 36.5   & 82.05 & 46.46 & 85.9 & 59.6& 84.03& 47.24 \\
% $\Delta$ & +14.1 & +16.8   & +3.30 & +12.61 & +22.3 & +17.2 & +3.30 & +12.74 \\
% \bottomrule
% \end{tabular}
% \end{table*}
\begin{table*}[htbp]
\centering
\caption{Summary evaluation and comparison of pass rate of different methods. For summary evaluation, we report BERTScore and SBERT. For decompilation quality, we report the percentage of outputs that can compile successfully and pass all test assertions.}
\label{tab:correctness_comparison}
\begin{tabular}{lcccccccc}
\toprule
& \multicolumn{4}{c}{HumanEval-X} & \multicolumn{4}{c}{MBXP} \\ 
\cmidrule(lr){2-5} \cmidrule(lr){6-9}
Method & Compile & Correct & BERTScore & SBERT & Compile & Correct & BERTScore & SBERT \\ 
\midrule
 & & &  & \textbf{GPT-3.5} & & & &  \\
\midrule
1-shot & 38.2 & 13.4 & 80.9   & 38.3 & 48.0& 19.6& 83.3& 36.3\\
3-shot  & 49.2 & 18.3 & 82.7   & 42.8 & 41.5 & 21.1 & 86.3 & 45.0\\
5-shot  & 50.7 & 22.1 & \textbf{84.5}   & 50.0 & 38.5 & 18.8 & \textbf{88.4} & \textbf{52.0}\\
10-shot & 51.6 & 23.4 & 84.1   & 50.7 & 23.9 & 10.2 & 86.9 & 50.6\\
StackSight & \textbf{58.1} & \textbf{31.4} & 83.6 & \textbf{51.8} & \textbf{54.9}& \textbf{29.5}& 84.6& 51.3 \\
\midrule
 & & &  & \textbf{Code Llama} & & & &  \\
\midrule
1-shot  & 47.7 & 13.5 &   79.5 & 37.1& 38.8& 10.3& 81.5& 37.0 \\
3-shot  & 42.6 & 10.8 & 79.9 & 40.4 & 21.2 & 5.5 & 82.1 & 40.9\\
5-shot  & 46.0 & 12.7 & 80.3 & 41.7 & 31.1 & 8.2 & 82.1 & 39.2\\
10-shot & 50.0 & 10.7 & 80.6 & 43.7 & 29.2 & 9.0 & 82.4 & 40.4\\
StackSight & \textbf{54.2} & \textbf{24.6}  & \textbf{83.1} & \textbf{53.5} & \textbf{41.8}& \textbf{15.4}& \textbf{85.4}& \textbf{55.0}\\
% $\Delta$ & +6.5 & +11.1   & +3.66 & +16.36 & +3.0& +5.1& +3.94 & +18.04 \\
\midrule
 & & &  & \textbf{GPT-4} & & & &  \\
\midrule
1-shot      & 46.5 & 19.7 & 78.75   & 33.9 & 63.6 & 42.4& 80.7& 34.5\\
StackSight  & 60.6 & 36.5   & 82.05 & 46.5 & 85.9 & 59.6& 84.0& 47.2 \\
$\Delta$ & +14.1 & +16.8   & +3.30 & +12.6 & +22.3 & +17.2 & +3.3 & +12.7 \\
\bottomrule
\end{tabular}
\end{table*}

\noindent \textbf{Discussion}\quad 
This correctness evaluation focuses on functional equivalence rather than syntactic similarity, which is a deliberate choice to match the intended use cases. In real use scenario, users are only presented with the WebAssembly code but not the original source code, and during compilation, identifiers and logic may change depending on the compiler, compilation flags, etc. Syntax may change and a different set of variables may be used, while the core functionality should be invariant. By using a test-driven approach, we can objectively measure the extent to which the decompiled code represents the functionality of the original WebAssembly. 

Among the five steps, the only manual intervention arises in Step 4 for fixing function argument pass-by-reference. We deliberately choose not to employ scripts to fix it automatically because altering function arguments may inadvertently modify function semantics. We manually analyze each case to ensure that adjustments preserve the original logic. 

\subsubsection{Results}
Our experiment results, detailed in Table~\ref{tab:correctness_comparison}, reveal that StackSight with one-shot still achieves the highest performance in the majority of the metrics, despite longer prompts and additional demonstration examples in the ICL baselines. 
These results demonstrate that our approach exhibits a superior ability to accurately capture and reflect the true purposes of the WebAssembly code, thus yielding more successful decompilation outcomes.

Regarding decompilation correctness, StackSight demonstrates the highest scores across all models and baselines.  
More specifically, we notice that StackSight brings twice as much improvement on HumanEval-X as on MBXP. Further analysis reveals that the discrepancy may be because many code snippets in MBXP are synthesized by a code generation model~\cite{mbxp_dataset}. These snippets, while passing test cases, often fail to align with the original functional objectives. Despite this challenge, StackSight still manages to achieve a 5 to 10 percentage increase in functional correctness. 
This indicates that StackSight significantly improves the interpretation of WebAssembly code, leading to more accurate decompilation. 
In addition, larger models (GPTs) benefit more from StackSight than Code Llama-7b, which is consistent with previous observations that complex reasoning abilities are more prominent in larger models ($\geq$ 100B)~\cite{wei2022emergent, fu2023specializing}.

StackSight also exhibits competitive performance in BERTScore and SBERT metrics for GPT-3.5 and highest performance for Code Llama. This suggests that StackSight-equipped LLMs excel at grasping the high-level functionalities of the code and generating natural language summaries that are semantically closer to the actual intentions of the original code.

In addition, we observe that increasing the number of demonstration examples improves the summarization and decompilation quality in most cases. This suggests that further enhancing StackSight's performance is possible by providing more, higher-quality examples. 

\subsubsection{Case Study}
We present two case studies that exemplify the effectiveness of StackSight.
The first case study, illustrated in Figure~\ref{fig:example}, showcases the capabilities of our static analysis algorithm. It accurately represents stack operations, delineates variable relationships, and navigates if-else branches. This detailed representation provides the LLM with rich contextual information, enabling it to successfully predict data types, restore variable semantics, succinctly summarize functional objectives, and generate a functionally equivalent version of the C++ code. 
%This example demonstrates how StackSight enhances the LLM's understanding and interpretation of WebAssembly code.

\begin{figure}[h]
    \centering
    % \vspace{-0.5em}
    \includegraphics[width=\columnwidth]{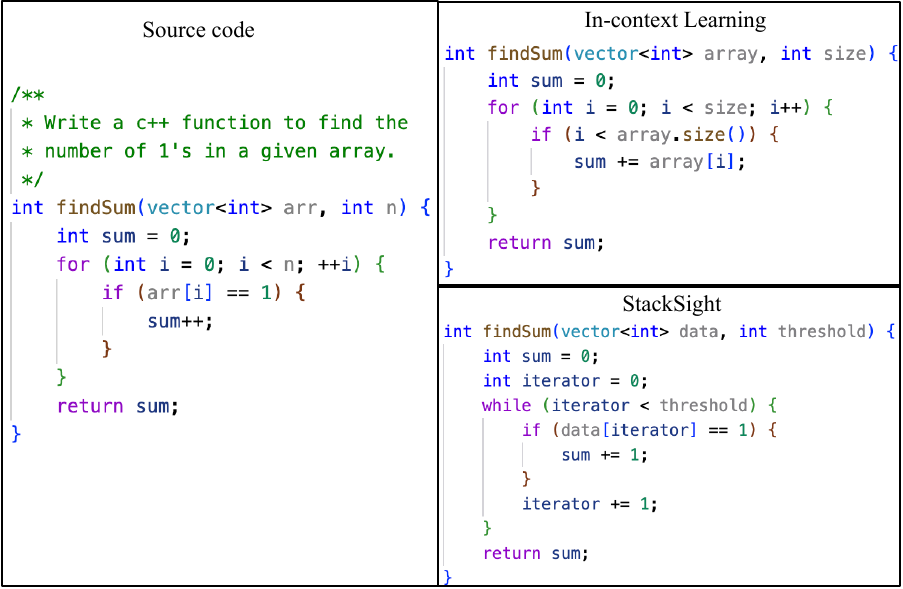}
    \caption{Hallucination case study.}
    \label{fig:case}
    %\vspace{-1em}
\end{figure}

The second case study, shown in Figure~\ref{fig:case}, focuses on StackSight's ability to counter hallucination, especially in scenarios with misleading function names. In this example, the function named \texttt{findSum} is actually intended to count the occurrence of 1's in a given array. We present two versions of decompiled code generated by gpt-3.5-turbo with ICL and with StackSight. The ICL version falsely generates a function that calculates the sum of all elements in the array, misled by the function name. In contrast, the StackSight version
accurately discerns the true purpose of the function. This case study highlights that StackSight helps elicit complex reasoning capabilities in LLMs and its potential to correct misinterpretations in decompilation tasks.

% \ww{This paragraph seems to repeat the previous paragraph.}
% The second case study, shown in Figure~\ref{fig:case}, focuses on StackSight's ability to mitigate hallucination effects, especially in scenarios with misleading function names. In this example, the function named "\textit{findSum}" is actually intended to count the occurrences of the number 1 in the given array. We compare two decompiled code versions: one generated by gpt-3.5-turbo using ICL and the other by StackSight. The ICL approach inaccurately produces a function that sums up all array elements, misled by the function's name. In contrast, StackSight's version accurately discerns the true purpose of the function. This case study highlights StackSight helps elicit the complex reasoning capabilities in LLMs and its potential to correct misinterpretations in decompilation tasks.

% \textbf{Failure case: summarize in general language. }: investigating instances where the pipeline fails or underperforms to identify causes and limitations of our approach

\subsection{RQ2: User Study}\label{sec:user-study}
To rigorously evaluate the efficacy of our decompiling methodology, we conduct a structured user study. We begin by randomly selecting twenty code snippets from each dataset for the user study. For every code snippet, we initiated the decompilation process utilizing two distinct approaches: the baseline In Context Learning (ICL) method and the StackSight approach, employing both Code Llama and GPT-3.5-turbo models. This procedure yielded four decompiled versions for each snippet. 

Subsequently, we carefully developed three multiple-choice questions per code snippet based on the source code, aligning with three distinct objectives: functionality comprehension, input-output relationship understanding, and edge case handling. These objectives range from a general grasp of the code's purpose to intricate details. Correct answers to these evaluation questions indicate that the decompiled code effectively captures and conveys the original functionality in a manner that is both accurate and human-readable. See~\ref{sec:user_study_detail} for more details.

% For each code snippet, we classify it alongside its counterpart from the same dataset, generated by the same model. This entails comparing one version decompiled using the StackSight approach with another produced via the baseline method. Subsequently, participants will conduct evaluations of these pairs in two distinct phases.

\textbf{Functional Comprehension}\quad
Participants are presented with three multiple-choice questions to answer based on the StackSight version and the baseline version. It aims to test how well participants can comprehend the given code snippet, thereby assessing whether our decompilation approach aids in producing code that is not only functionally equivalent but also easily comprehensible.

\textbf{Code Similarity}\quad
Following the initial comprehension evaluation, participants are presented with the original source code and its two decompiled versions. They are tasked with choosing which decompiled version most closely resembles the semantic content of the original source code, through comparison. This stands in parallel to the test-driven approach, as the decompiled code may capture the essence of the binary code but fails to pass the test cases due to some edge case handling. This process provides an additional layer of evaluation as to how well StackSight manages to recover high-level code semantics.

\subsubsection{User Study Statistics}
We recruited 15 participants in our user study. To better mimic a real-world scenario, our participants are either graduate students in engineering or professional developers. All participants are proficient in C/C++.
The total time for completing our questions ranges from 45 to 60 minutes. We ensure that for each designed question, there are two to three responses from the participants to reduce personal bias.

% \ww{Any brief description of the participants? e.g., number of people, background (e.g., major, programming skills, graduate/undergraduate, male/female?, etc.). Time taken to complete the study.}

\begin{figure}[h]
    \centering
    %\vspace{-0.5em}
    \includegraphics[width=\columnwidth]{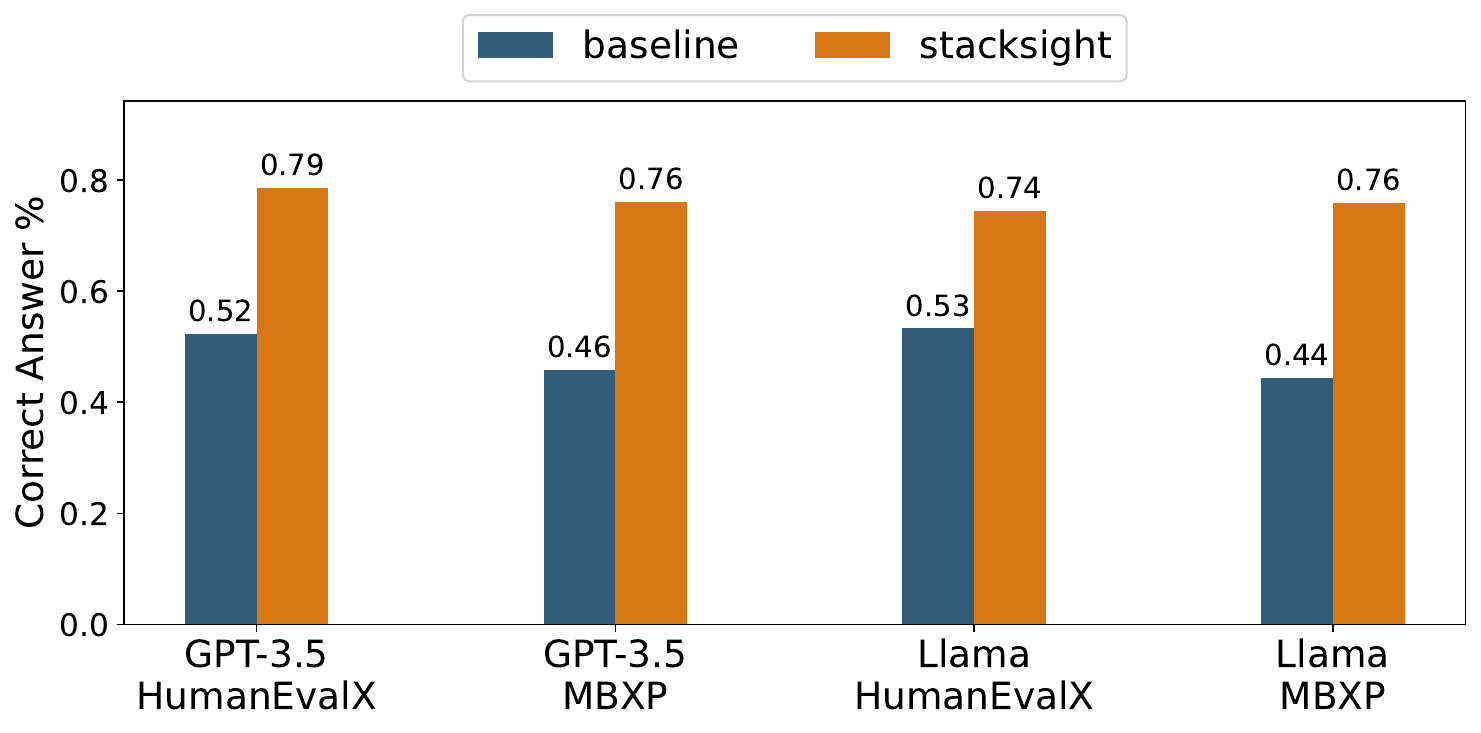}
    \caption{Comparison of correct answer percentages.}
    \label{fig:userstudy_functionality}
    %\vspace{-1em}
\end{figure}
\begin{figure}[h]
    \centering
    %\vspace{-0.5em}
    \includegraphics[width=\columnwidth]{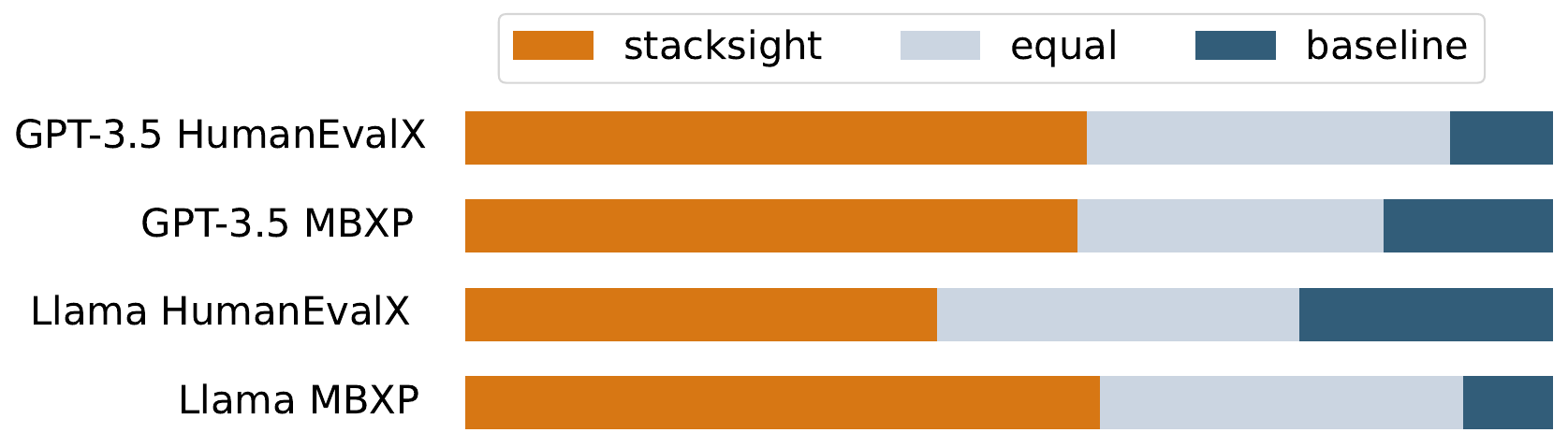}
    \caption{Comparison of similarity ratios.}
    \label{fig:userstudy_similarity}
    %\vspace{-1em}
\end{figure}

% \begin{table}[htbp]
% \centering
% \caption{User Study Results: We report the average percent of correct answers for the multiple-choice questions from users and the average percent of the times that the version is believed to more similar to the source codes. \ww{Is this table intended to present the same results as Figures 4 and 5?}}
% \label{tab:user_study}
% \begin{tabular}{lcccc}
% \toprule
% & \multicolumn{2}{c}{HumanEval-X} & \multicolumn{2}{c}{MBXP} \\ 
% \cmidrule(lr){2-3} \cmidrule(lr){4-5}
% Metric & Correct & Similar & Correct & Similar\\ 
% \midrule
% GPT-3.5  & 52.3 & 9.5 & 45.8 & 15.6  \\
% + StackSight  & 78.5 & 57.1 & 76.0 & 56.2 \\
% \midrule
% Code Llama  & 53.3 & 23.3 & 44.4 & 8.3 \\
% + StackSight & 74.4 & 43.3& 75.9 & 58.3 \\
% \bottomrule
% \end{tabular}
% \end{table}

Figure \ref{fig:userstudy_functionality} presents the results of our study on functionality comprehension. 
We report the average percentage of correct responses from participants given the decompiled code. Models equipped with StackSight markedly outperform the baseline, achieving at least a 20\% higher rate of correct answers. Figure \ref{fig:userstudy_similarity} depicts the results of our code similarity evaluation, where each bar represents the proportion of participants who identified code snippets as closely resembling the source code. Remarkably, StackSight approach exhibited a higher resemblance ratio in comparison to the baseline. 

%The correct answer rate seems pretty close across different models and datasets, but one observation is that StackSight yields a slightly greater increase in accuracy on the MBXP dataset compared to HumanEval-X. Through case studies, we found that this discrepancy is often due to a misalignment between the actual intent of the code and the function names in the MBXP dataset, leading to inaccuracies when the LLM model generates code without additional context, However, with StackSight, the LLM models are furnished with additional context, enabling them to generate responses with greater precision. 

%It is noteworthy that the performance of two approaches on the Llama model with the HumanEval-X dataset appears very close, aligning with the understanding of functionality where users achieved the highest percentage of correct answers using Llama and HumanEval-X.

\subsection{RQ3: Ablation Study} 
To understand the contribution of each component within StackSight, we conduct an ablation study on the following elements: stack visualization, recovery of variable semantics, and generation of natural language summaries. All prompts can be found in Appendix~\ref{app:ablation_prompts}.

\subsubsection{Results}
\begin{table}[h]
\centering
\caption{Ablation study.}
\label{tab:ablation_study}
\begin{tabular}{l c c }
\toprule
\% Correct & HumanEval-X & MBXP  \\ \midrule
GPT-3.5     & 13.4 & 19.6\\
+StackSight & 31.4 & 29.5\\ \midrule
%No stack visualization & 22.0 & 25.0  \\
No stack visualization & 23.6 & 26.5  \\
Basic CoT  & 22.2 & 26.2  \\
No variable semantics & 24.8 & 22.7  \\
No function summary  & 20.7 & 24.3  \\
\bottomrule
\end{tabular}
\end{table}
As detailed in Table~\ref{tab:ablation_study}, we report the percentage of decompiled code snippets that can pass all test cases by removing each component for the two datasets. 
Given that the annotated binary code is nearly twice as large as the original version, we include another example for ``No stack visualization" ablation to match the prompt size. 

Removing either the explicit stack visualization or the variable semantics recovery significantly reduces the contextual information, leading to a degradation of decompilation accuracy. The removal of stack visualization leads to a decrease in the successful decompilation rate by $7.8\%$ and $3\%$ for the two datasets, and the absence of variable semantics recovery results in a performance drop by $6.6\%$ and $6.8\%$. These results underscore the importance of translating low-level WebAssembly operations and data types to high-level counterparts. 

The omission of the natural language summary component also significantly impacts performance, with a decrease of $10.7\%$ and $5.2\%$. Removing the step of summarization causes LLM to resort to a more literal transcription of the WebAssembly code, resulting in verbatim output and incorrect C++ code. These findings underscore the integral role each component plays in our decompilation pipeline.

We also include the basic chain-of-thought prompting by using a short system prompt, while keeping everything else the same as the ``No stack visualization" ablation. We observe that the extended system prompt leads to marginal improvement of $1.4\%$ and $0.3\%$.

%\subsection{RQ4: Real-world Case Studies} 
% Upon completing our pipeline, we plan to conduct case studies focusing on:
% \begin{itemize}
%     \item Multiple Functions: To better simulate real-world scenarios, we have expanded the two datasets to include files with multiple functions, where functions may interdependently call each other. Additionally, we have written the test cases and function summaries to cover these additional functions, ensuring a comprehensive evaluation of their functionality and interactions. 
%     \item Real-world project applicability: testing the pipeline's applicability in larger, real-world WebAssembly projects.
%     \item Malicious code detection: exploring the pipeline's utility in identifying WebAssembly malicious code, especially as malicious logic written in WebAssembly is hard to detect. This study aims to test the usefulness of our approach in malware detection in web applications. 
% \end{itemize}
%\subsubsection{Results}
\section{Conclusion}
In this paper, we propose StackSight, a three-stage neural-symbolic method that decompiles WebAssembly into C++ code. Each stage of our pipeline is carefully designed to tackle a challenge in understanding WebAssembly. We have shown empirically that StackSight significantly enhances the functional correctness of decompiled C++ snippets and improves the semantic relevance of model generated summaries. We have also conducted case studies to showcase how StackSight mitigates the risks of misinterpretation by LLMs in ambiguous contexts.
Furthermore, the notably higher win-rate of StackSight generated C++ code in our user study further underscores the effectiveness of StackSight. Our work pushes forward the capabilities of current WebAssembly decompilation toolkits. Hopefully, our work can shed light upon eliciting the power of large language models for low-resource code languages.

\section*{Acknowledgements}

% If a paper is accepted, the final camera-ready version can (and
% usually should) include acknowledgements.  Such acknowledgements
% should be placed at the end of the section, in an unnumbered section
% that does not count towards the paper page limit. Typically, this will 
% include thanks to reviewers who gave useful comments, to colleagues 
% who contributed to the ideas, and to funding agencies and corporate 
% sponsors that provided financial support.

We thank the anonymous reviewers for their constructive feedback.
This work was partially supported by the US National Science
Foundation under Grant No. 2321444. Any opinions, findings, and
conclusions in this paper are those of the authors only and do not
necessarily reflect the views of our sponsors.

\section*{Impact Statement}
Our work will enhance the comprehension of WebAssembly code distributed on the Internet. 
Binary codes are frequently minified or obfuscated, and they may serve unintended or malicious functional purposes. Our work proposes StackSight, which proves extremely useful by decompiling these binaries into high-level, human-readable programs complemented by natural language summaries.
We hope that our work can reduce the prevalence of WebAssembly-based malware, fostering a more transparent environment for third-party WebAssembly code.

\bibliographystyle{icml2024}
\bibliography{references}

%%%%%%%%%%%%%%%%%%%%%%%%%%%%%%%%%%%%%%%%%%%%%%%%%%%%%%%%%%%%%%%%%%%%%%%%%%%%%%%
%%%%%%%%%%%%%%%%%%%%%%%%%%%%%%%%%%%%%%%%%%%%%%%%%%%%%%%%%%%%%%%%%%%%%%%%%%%%%%%
% APPENDIX
%%%%%%%%%%%%%%%%%%%%%%%%%%%%%%%%%%%%%%%%%%%%%%%%%%%%%%%%%%%%%%%%%%%%%%%%%%%%%%%
%%%%%%%%%%%%%%%%%%%%%%%%%%%%%%%%%%%%%%%%%%%%%%%%%%%%%%%%%%%%%%%%%%%%%%%%%%%%%%%
\newpage
\appendix
% \onecolumn
\appendix
\newpage
\section*{Appendix}

\section{Original Dataset Examples}\label{app:dataset_examples}
% \noindent \textbf{HumanEval-X}
\subsection{HumanEval-X Example}\label{app:humanevalx_example}

\begin{lstlisting}[style=customcpp]
// prompt:
/*
Check if in given vector of numbers, are any two numbers closer to each other than
given threshold.
>>> has_close_elements({1.0, 2.0, 3.0}, 0.5)
false
>>> has_close_elements({1.0, 2.8, 3.0, 4.0, 5.0, 2.0}, 0.3)
true
*/
#include<stdio.h>
#include<vector>
#include<math.h>
using namespace std;
bool has_close_elements(vector<float> numbers, float threshold){
// canonical solution:
    int i,j;
    
    for (i=0;i<numbers.size();i++)
    for (j=i+1;j<numbers.size();j++)
    if (abs(numbers[i]-numbers[j])<threshold)
    return true;

    return false;
}
// test:
#undef NDEBUG
#include<assert.h>
int main(){
    vector<float> a={1.0, 2.0, 3.9, 4.0, 5.0, 2.2};
    assert (has_close_elements(a, 0.3)==true);
    assert (has_close_elements(a, 0.05) == false);

    assert (has_close_elements({1.0, 2.0, 5.9, 4.0, 5.0}, 0.95) == true);
    assert (has_close_elements({1.0, 2.0, 5.9, 4.0, 5.0}, 0.8) ==false);
    assert (has_close_elements({1.0, 2.0, 3.0, 4.0, 5.0}, 2.0) == true);
    assert (has_close_elements({1.1, 2.2, 3.1, 4.1, 5.1}, 1.0) == true);
    assert (has_close_elements({1.1, 2.2, 3.1, 4.1, 5.1}, 0.5) == false);
    
}
\end{lstlisting}

% \noindent \textbf{MBXP}
\newpage
\subsection{MBXP Example}\label{app:mbxp_example}

\begin{lstlisting}[style=customcpp]
// prompt:
#include <bits/stdc++.h>
using namespace std;


/**
 * Write a c++ function to check whether the two numbers differ at one bit position only or not.
 * > differAtOneBitPos(13, 9)
 * true
 * > differAtOneBitPos(15, 8)
 * false
 * > differAtOneBitPos(2, 4)
 * false
 */
bool differAtOneBitPos(int a, int b) {
// canonical solution:
    return (a == b) || (a == 13 && b == 9);
}

// test:
template <typename T> bool compare(T a, T b){ 
    return a == b; 
}

int main(int argc, char* argv[]) {
    bool x0 = differAtOneBitPos(13, 9);
    if (!(compare(x0, true))) {
        throw runtime_error("Exception -- test case 0 did not pass.");
    }

    bool x1 = differAtOneBitPos(15, 8);
    if (!(compare(x1, false))) {
        throw runtime_error("Exception -- test case 1 did not pass.");
    }

    bool x2 = differAtOneBitPos(2, 4);
    if (!(compare(x2, false))) {
        throw runtime_error("Exception -- test case 2 did not pass.");
    }

    return 0;
}
\end{lstlisting}
\section{Prompts}\label{appendix:prompt_eg}
\subsection{In-Context Learning Prompt}\label{in_context_prompt_eg}
\begin{lstlisting}[breaklines=true, breakatwhitespace=false, basicstyle=\ttfamily,]
System: You are a professional developer who translates WebAssembly text into C++ code.
Given the following WebAssembly text file, please provide the equivalent C++ code.
Human: 
\end{lstlisting}
\begin{lstlisting}[style=customwat]
(module
  (type (;0;) (func))
  (type (;1;) (func (param i32) (result i32)))
  (func (;0;) (type 0))
  (func (;1;) (type 1) (param i32) (result i32)
    local.get 0
    local.get 0
    i32.mul)
  (export "__wasm_call_ctors" (func 0))
  (export "__wasm_apply_data_relocs" (func 0))
  (export "_Z18car_race_collisioni" (func 1)))
\end{lstlisting}
\begin{lstlisting}[breaklines=true, breakatwhitespace=false, basicstyle=\ttfamily,]
AI: ```cpp
\end{lstlisting}
\begin{lstlisting}[style=customcpp]
#include<stdio.h>
#include<math.h>
using namespace std;
#include<algorithm>
#include<stdlib.h>
int car_race_collision(int n){
    return n*n;
}
\end{lstlisting}
\begin{lstlisting}[breaklines=true, breakatwhitespace=false, basicstyle=\ttfamily,]
```
Human: 
\end{lstlisting}
\begin{lstlisting}[style=customwat]
(module
(type (;0;) (func))
(type (;1;) (func (param i32) (result i32)))
(func (;0;) (type 0))
(func (;1;) (type 1) (param i32) (result i32)
    local.get 0
    i32.const 1
    i32.add
    local.get 0
    i32.mul
    i32.const 2
    i32.div_s)
(export "__wasm_call_ctors" (func 0))
(export "__wasm_apply_data_relocs" (func 0))
(export "_Z8sum_to_ni" (func 1)))
\end{lstlisting}

\newpage
\subsection{CoT Learning Prompt}
\label{app:cot_prompt}
\begin{lstlisting}[breaklines=true, breakatwhitespace=false, basicstyle=\ttfamily,]
System: You are a professional reverse engineer tasked with translating WebAssembly text (WAT) into C++ code. The process involves interpreting annotated WAT code, understanding its stack state and conditional branches, predicting data types and semantics, summarizing in natural language, and finally generating equivalent C++ code. Let's do it step by step.

Step 1: Data Type Prediction
Given annotated WAT code, examine each line's comments that visualize the virtual stack state.
Identify and predict the high-level C/C++ data types for input parameters and return values based on their operations and usage in the code.

Step 2: Semantic Variable Recovery
Analyze stack visualizations to extract variable relationships.Recover and describe the semantic meaning of each variable, including parameters, returns, and local variables. Use 'l0', 'l1', etc., to represent variables corresponding to local.get 0, local.get 1, and so on.

Step 3: Function Summary
Revisit the source WAT function based on your understanding of variable semantics. Summarize the function's purpose and operation in a few lines, focusing on the overall functionality.

Step 4: C++ Code Generation
Going back to the original WAT code, decompile it into equivalent C++ code based on the variable semantics and function summary. Write the C++ code in a standard format, avoiding direct line-by-line translation from WAT. Enclose C++ code within "cpp" and "".

Format for Output:
Clearly indicate each step's output with "Step 1:", "Step 2:", etc.
Provide explanations and reasoning for each step's conclusion."
Human: 
\end{lstlisting}
\begin{lstlisting}[style=customwat]
(module $code_3.wasm
  (type (;0;) (func))
  (type (;1;) (func (param i32) (result i32)))
  (import "env" "memory" (memory (;0;) 0))
  (func $__wasm_call_ctors (type 0))
  (func $below_zero_std::__2::vector<int__std::__2::allocator<int>>_ (type 1) (param i32) (result i32)
    (local i32 i32 i32 i32)  ;; stack: []
    local.get 0  ;; [l0]
    i32.load offset=4  ;; [(l0[4])_mem]
    local.get 0  ;; [(l0[4])_mem, l0]
    i32.load  ;; [(l0[4])_mem, (l0)_mem]
    local.tee 4  ;; [(l0[4])_mem, l4] ;; l4 = (l0)_mem
    i32.sub  ;; [(l0[4])_mem-l4]
    i32.const 2  ;; [(l0[4])_mem-l4, 2]
    i32.shr_s  ;; [(l0[4])_mem-l4/4]
    local.set 1  ;; [] ;; l1 = (l0[4])_mem-l4/4
    loop  ;; label = @1
      block  ;; label = @2
        local.get 1  ;; [l1]
        local.get 3  ;; [l1, l3]
        local.tee 0  ;; [l1, l0] ;; l0 = l3
        i32.eq  ;; [l1==l0]
        if  ;; label = @3  ;; [] branch starts 
          local.get 1  ;; (c0) [l1] ;; c0: l1==l0
          local.set 0  ;; (c0) [] ;; l0 = l1
          br 1 (;@2;)  ;; (c0) [] break out of the block at @2
        end
        local.get 0  ;; (c1) [l0] ;; c1: !(l1==l0)
        i32.const 1  ;; (c1) [l0, 1]
        i32.add  ;; (c1) [l0+1]
        local.set 3  ;; (c1) [] ;; l3 = l0+1
        local.get 4  ;; (c1) [l4]
        local.get 0  ;; (c1) [l4, l0]
        i32.const 2  ;; (c1) [l4, l0, 2]
        i32.shl  ;; (c1) [l4, l0*4]
        i32.add  ;; (c1) [l4+l0*4]
        i32.load  ;; (c1) [(l4+l0*4)_mem]
        local.get 2  ;; (c1) [(l4+l0*4)_mem, l2]
        i32.add  ;; (c1) [(l4+l0*4)_mem+l2]
        local.tee 2  ;; (c1) [l2] ;; l2 = (l4+l0*4)_mem+l2
        i32.const 0  ;; (c1) [l2, 0]
        i32.ge_s  ;; (c1) [l2>=0]
        br_if 1 (;@1;)  ;; (c2) [] continue the loop at @1 OR (c3) [] move on to next line ;; c2: c1 && l2>=0 ;; c3: c1 && !(l2>=0)
      end
    end
    local.get 0  ;; [l0]
    local.get 1  ;; [l0, l1]
    i32.lt_u)  ;; [l0<l1]
  (export "__wasm_call_ctors" (func $__wasm_call_ctors))
  (export "__wasm_apply_data_relocs" (func $__wasm_call_ctors))
  (export "_Z10below_zeroNSt3__26vectorIiNS_
  9allocatorIiEEEE" (func $below_zero_std::__2::vector<int__std::__2::allocator<int>>_)))
;; condition dict
;; c0: l1==l0
;; c1: !(l1==l0)
;; c2: c1 && l2>=0
;; c3: c1 && !(l2>=0)
\end{lstlisting}
\begin{lstlisting}[breaklines=true, breakatwhitespace=false, basicstyle=\ttfamily,]
AI: 
Step 1:Data Type Prediction

The function's parameter is used in memory operations with an offset (i32.load offset=4), suggesting it's likely a pointer to an array or a complex data structure. Shift operations (i32.shr_s / i32.shl) by 2 places are observed, so it is likely pointer arithmetic of an array of 4-byte objects, such as an integer array. The return value is the result of a comparison.Thus, the return type is likely a boolean (True/False) in C++. 


Step 2: Semantic Variable Recovery

l0: Initially loaded from memory and later used as a loop counter. It likely represents a pointer to the start of an array or a similar structure, based on memory access patterns (l4 = l0_mem). As it's incremented in the loop (l0=l3 and l3 = l0+1), it also serves as an index or counter within the array.

l1: Calculated as a shifted and subtracted value from l0, likely representing a boundary or limit within the data structure, possibly related to the size or an upper limit of an array (l1 = l0[4]_mem - l4 / 4 where l4 = l0_mem)

l2: Appears to accumulate values in a loop, suggesting it's used for a running total or sum. The way it's updated in each iteration indicates an accumulative operation (l2 = (l4+l0<<2_mem)+l2)

l3: Incremented in the loop, closely following l0, it seems to function as an auxiliary counter or a modified index within the loop (l3 = l0 + 1)

l4: Loaded directly from memory (l0_mem) and used in arithmetic operations related to other variables. It likely represents a base address or a reference point within the data structure, possibly the starting address of the array's data.


Step 3: Function Summary

The function iteratively processes elements of an integer array or vector. It involves accumulating values and checking if this cumulative sum falls below zero. The function terminates the iteration early if this condition is met, which is a typical pattern for functions that perform a check or calculation until a certain condition is no longer satisfied when the sum l2 is no longer positive The final comparison (i32.lt_u), comparing l0 and l1, suggests a boundary check, likely ensuring the loop does not exceed the array's limits.
In essence, this function appears to be designed to check whether the cumulative sum of elements in an integer array ever becomes negative. If so, it returns true, indicating this condition has been met. Otherwise, the function returns false.

Step 4: C++ Code Generation

```cpp
\end{lstlisting}
\begin{lstlisting}[style=customcpp]
\end{lstlisting}

\begin{lstlisting}[style=customcpp]
#include<stdio.h>
#include<vector>
using namespace std;
bool below_zero(vector<int> operations){
    int num=0;
    for (int i=0;i<operations.size();i++)
    {
        num+=operations[i];
        if (num<0) return true;
    }
    return false;
}
\end{lstlisting}
\begin{lstlisting}[breaklines=true, breakatwhitespace=false, basicstyle=\ttfamily,]
```
Human: 
\end{lstlisting}
\begin{lstlisting}[style=customwat]
(module $code_2.wasm
  (type (;0;) (func))
  (type (;1;) (func (param f32) (result f32)))
  (func $__wasm_call_ctors (type 0))
  (func $truncate_number_float_ (type 1) (param f32) (result f32)
    local.get 0  ;; [l0]
    block (result i32)  ;; label = @1
      local.get 0  ;; [l0, l0]
      f32.abs  ;; [l0, abs(l0)]
      f32.const 0x1p+31 (;=2.14748e+09;)  ;; [l0, abs(l0), 0x1p+31]
      f32.lt  ;; [l0, abs(l0)<0x1p+31]
      if  ;; label = @2  ;; [l0]
        local.get 0  ;; [l0, l0]
        i32.trunc_f32_s  ;; [l0, trunc(l0)]
        br 1 (;@1;)  ;; [l0] Break Out of The Loop at @1
      end
      i32.const -2147483648  ;; [l0, -2147483648]
    end
    f32.convert_i32_s  ;; [l0, f32(trunc(l0))]
    f32.sub)  ;; [l0-f32(trunc(l0))]
  (export "__wasm_call_ctors" (func $__wasm_call_ctors))
  (export "__wasm_apply_data_relocs" (func $__wasm_call_ctors))
  (export "_Z15truncate_numberf" (func $truncate_number_float_)))

\end{lstlisting}

\subsection{Ablation Study Prompts}
\label{app:ablation_prompts}

\subsubsection{No Stack Visualization}
We remove static analysis annotation from each of the Wasm binaries, while everything else remains the same as Appendix~\ref{app:cot_prompt}.
\subsubsection{No Stack Visualization + Basic CoT}
We replace the extended system prompt with a short system prompt, while keeping everything else the same as the ``No stack visualization” ablation. 

\begin{lstlisting}[breaklines=true, breakatwhitespace=false, basicstyle=\ttfamily,]
System: You are a professional reverse engineer tasked with translating WebAssembly text (WAT) into C++ code. Let's do it step by step.
\end{lstlisting}
\subsubsection{No Variable Semantics}
We remove Step 2 from the CoT learning prompt in Appendix~\ref{app:cot_prompt}, resulting in a decrease of 70 tokens in the prompt, which is negligible.
\subsubsection{No Function Summary}
We remove Step 3 from the CoT learning prompt in Appendix C.2, which leads to a decrease of 43 tokens in the prompt.

\section{Compilation Command}\label{compilation_cmd}
\begin{lstlisting}[style=custombash]
emcc "$target_cpp" -o "$output_wasm" \
    --profiling-funcs \ # keep function 
    # names
    -Wl,--demangle \ 
    -s SIDE_MODULE=1 \ # compile as 
    # side module
    -Oz # optimize size
wasm2wat "$output_wasm" -o "$output_wat"
\end{lstlisting}

\subsection{User Study}\label{sec:user_study_detail}
This section outlines the structure of our user study, describing the initial setup followed by an exemplification of a subset of the study.
\subsection{Setup}
Our study is conducted online, with a web app we developed to ensure a user-friendly and streamlined process. All data was collected anonymously to maintain participant privacy. Participants were first acquainted with the study through instructions and a visual tutorial within the application, which provided a clear understanding of the task requirements.
\subsection{A Sample of the study}
\begin{figure}[H]
    \centering
    \includegraphics[width=.95\columnwidth]{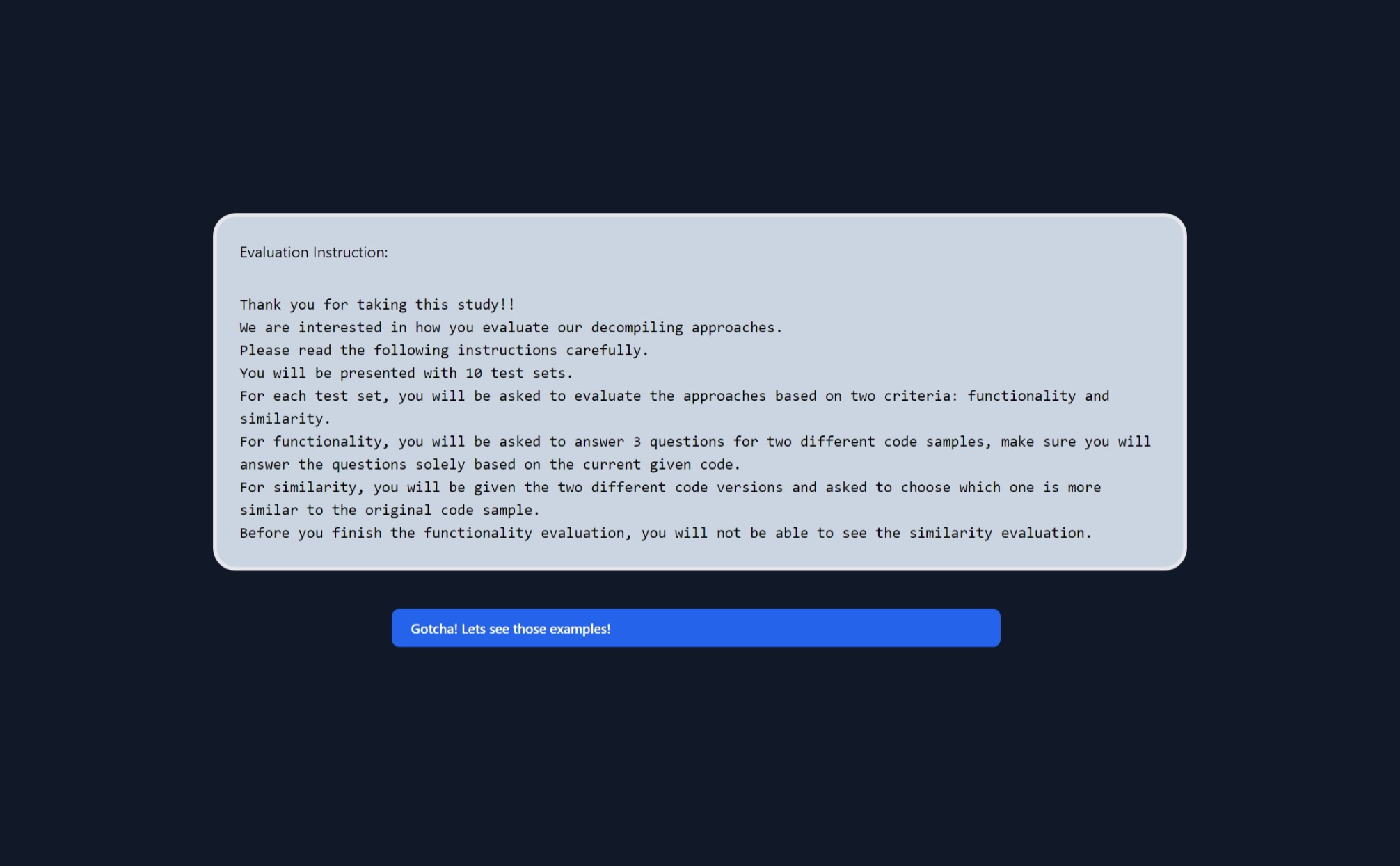}
    \caption{Task instructions.}
    \label{fig:system_instructions}
\end{figure}

\begin{figure}[H]
    \centering
    \includegraphics[width=.95\columnwidth]{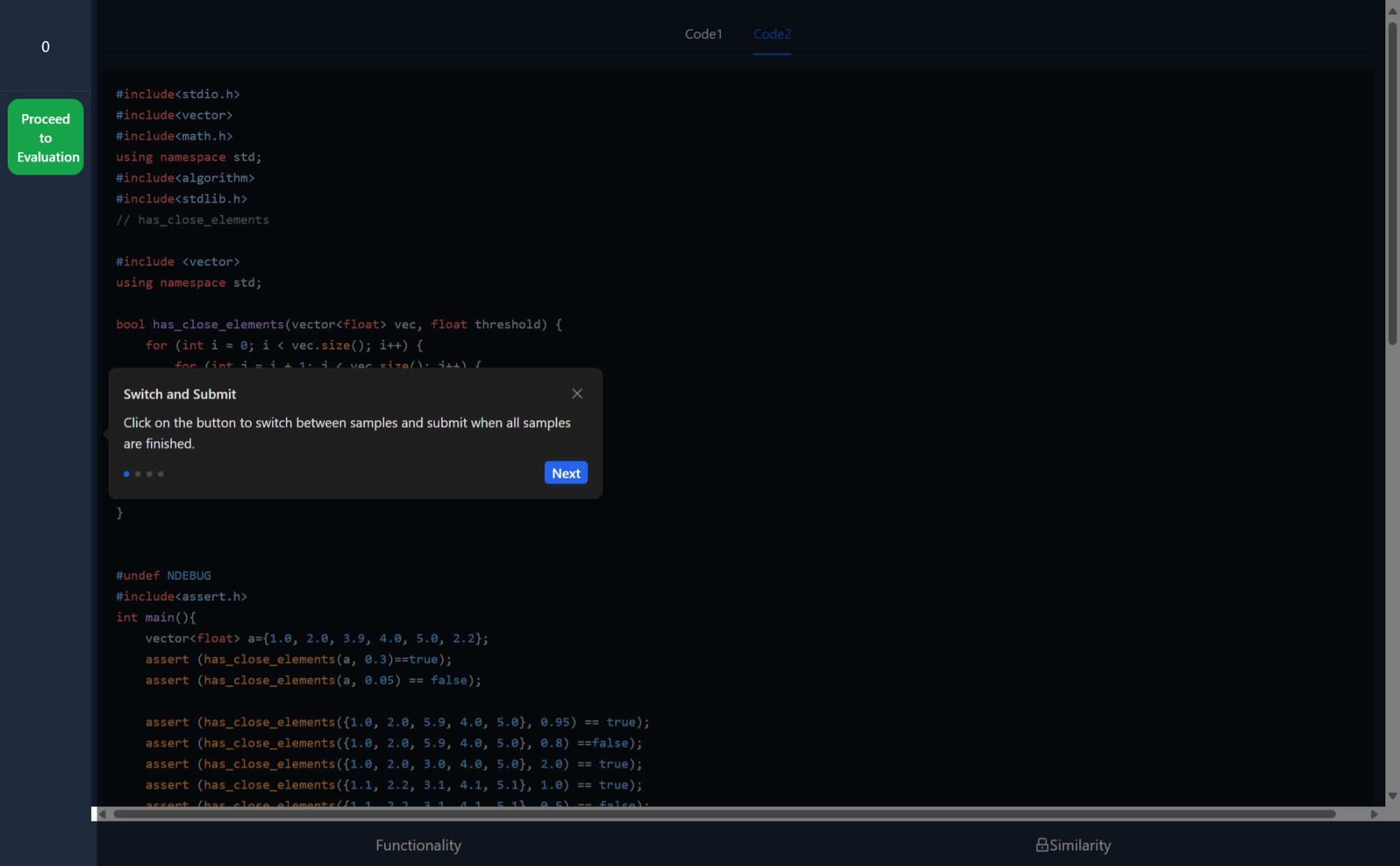}
    \caption{System tutorial sample 1.}
    \label{fig:system_tutorial_1}
\end{figure}

\begin{figure}[H]
    \centering
    \includegraphics[width=.9\columnwidth]{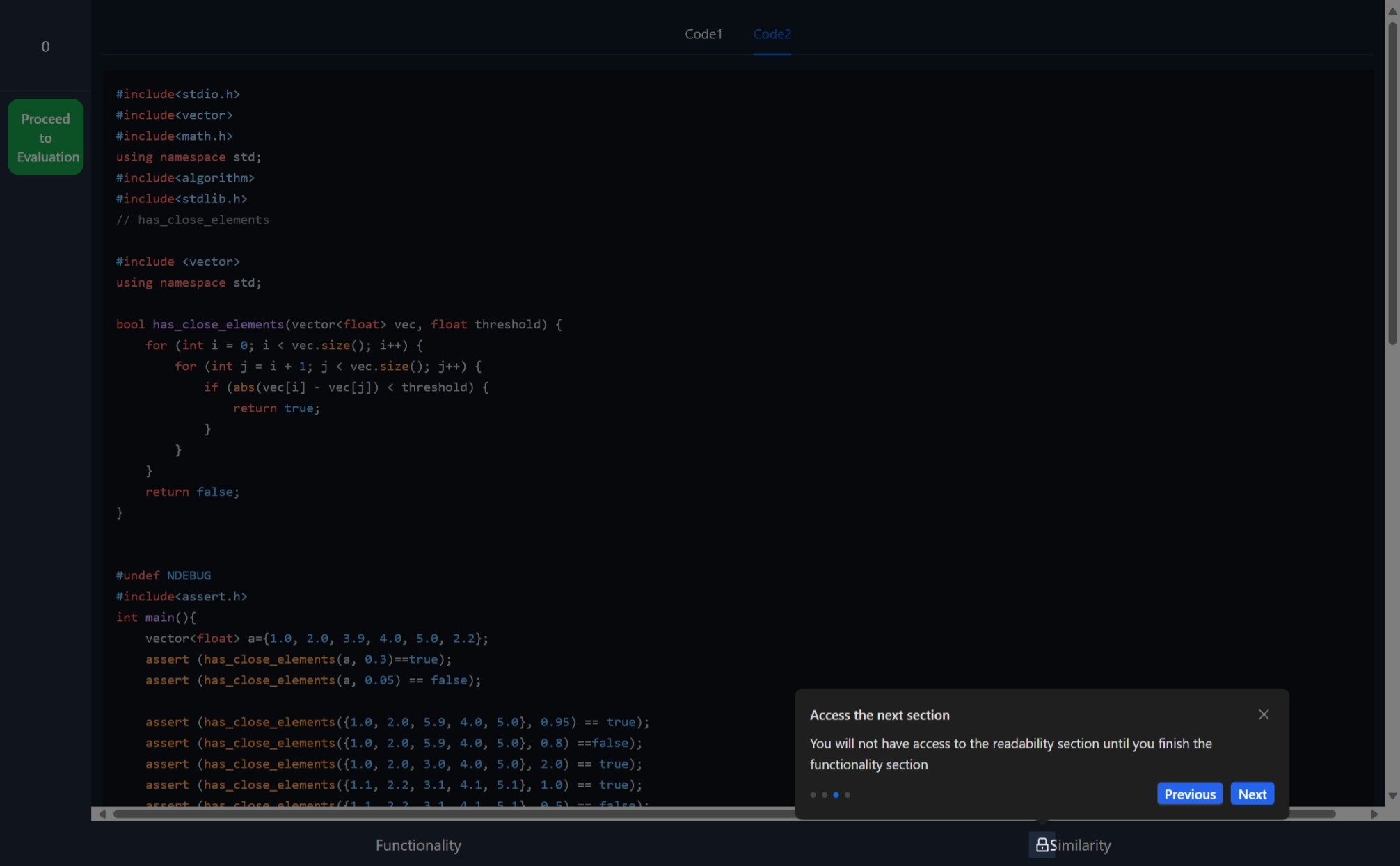}
    \caption{System tutorial sample 2.}
    \label{fig:system_tutorial_2}
\end{figure}

As depicted in Figure  \ref{fig:system_instructions}. Upon entering the evaluation system, participants will see the instructions first.
Then by clicking the button below, they will enter example page and they will see a example set with detailed tutorials of the complete user study process. The tutorials is consisted of many steps, for simplicity we show the tutorial with Figures \ref{fig:system_tutorial_1} and \ref{fig:system_tutorial_2} illustrating a selection of these.
This stepwise tutorial ensures that participants are well-informed about the study's progression, emphasizing that they must complete the functionality assessment before proceeding, thereby maintaining consistency throughout the study.

Next, we will show how we display our evaluation tasks.

\textbf{Functional Comprehension}
\begin{figure}[H]
    \centering
    \includegraphics[width=.9\columnwidth]{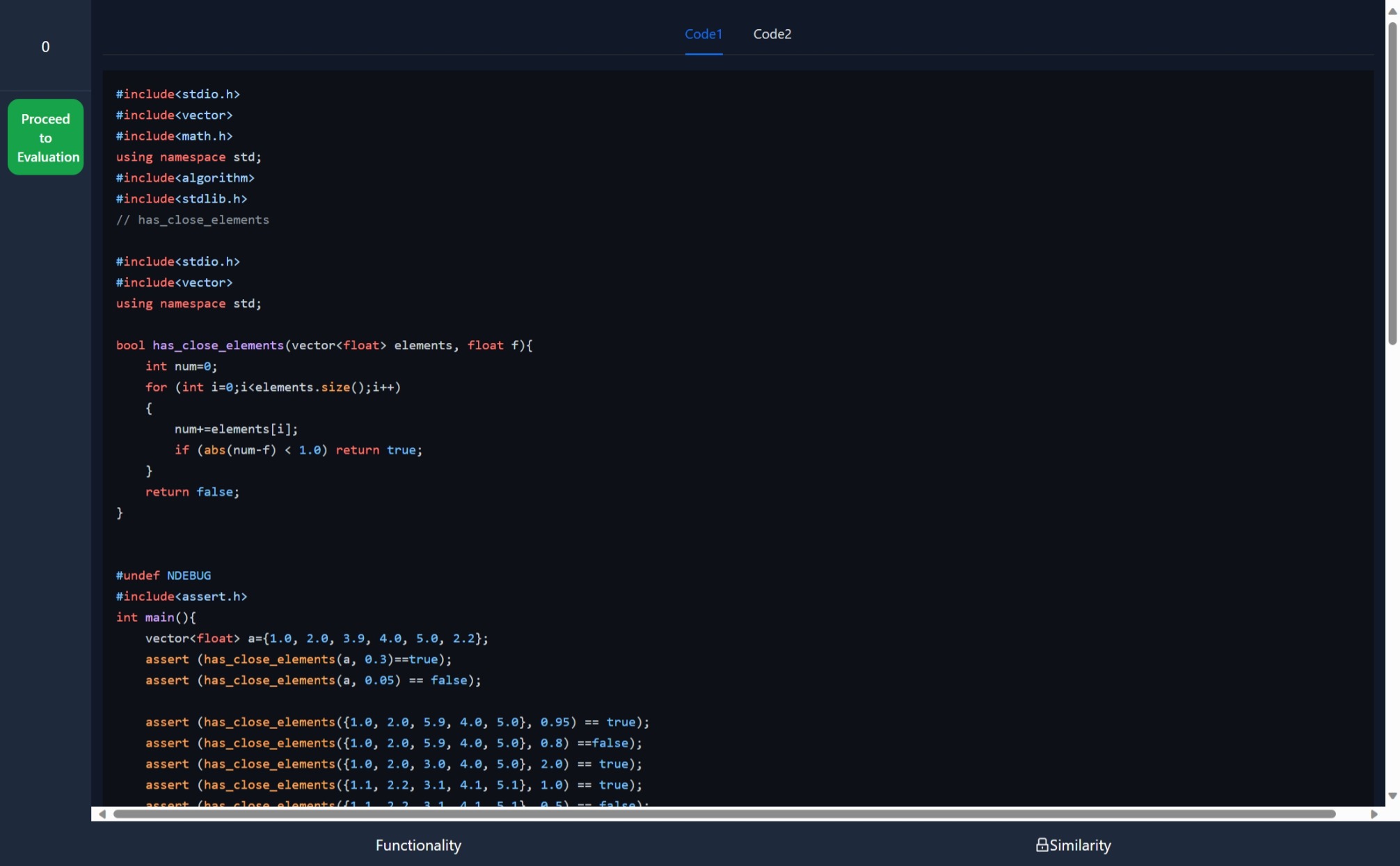}
    \caption{Functionality samples.}
    \label{fig:functionality_samples}
\end{figure}
\begin{figure}[H]
    \centering
    \includegraphics[width=.85\columnwidth]{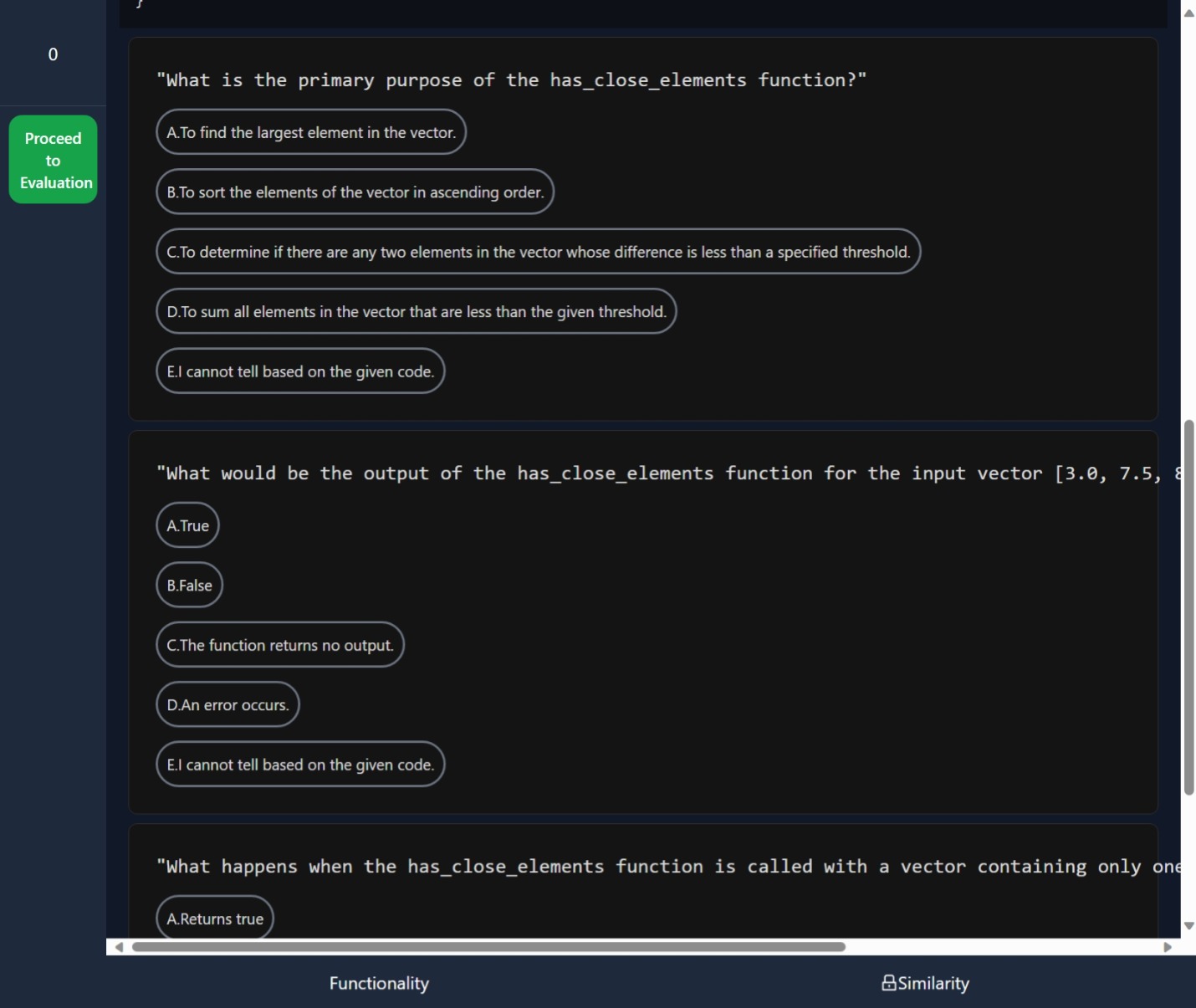}
    \caption{Functionality questions.}
    \label{fig:functionality_questions}
\end{figure}
In figure \ref{fig:functionality_samples} we can clearly see each set has two code samples—one derived from the baseline approach and the other from our StackSight approach. Accompanying these code samples, as presented in Figure \ref{fig:functionality_questions}, are three multiple-choice questions created based on source code. To address potential semantic variations between the source and generated codes, we include the option "I cannot tell based on the given code.". Participants' progress is visually restricted; a lock icon on the similarity section tab indicates that they must answer all questions before accessing the subsequent section.

\textbf{Code Similarity}
\begin{figure}[H]
    \centering
    \includegraphics[width=\columnwidth]{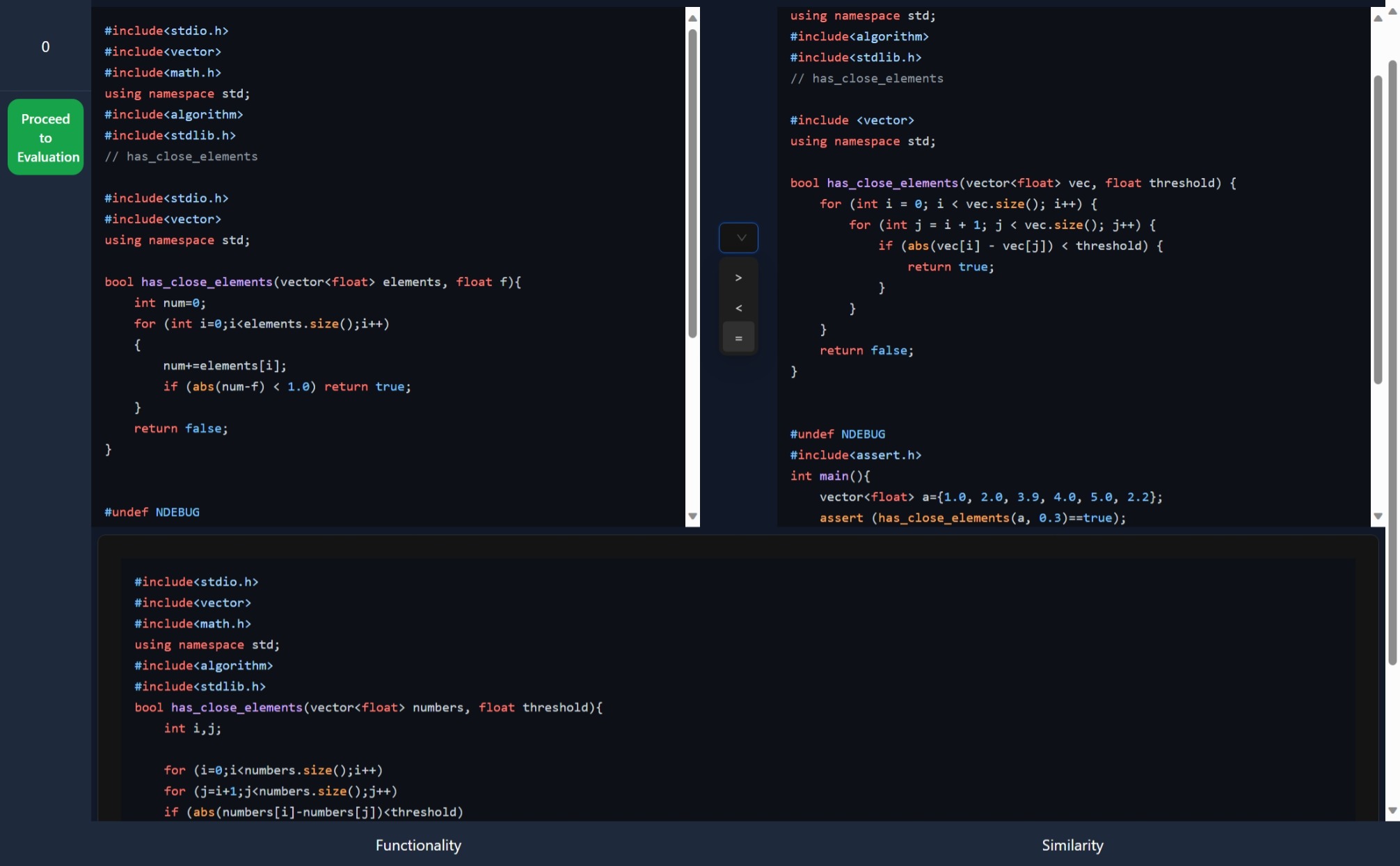}
    \caption{Similarity sample.}
    \label{fig:similarity_sample}
\end{figure}
Figure \ref{fig:similarity_sample} shows the section where participants compare two decompiled versions of code with the original, judging which decompiled code resembles the source more closely.
\begin{figure}[H]
    \centering
    \includegraphics[width=\columnwidth]{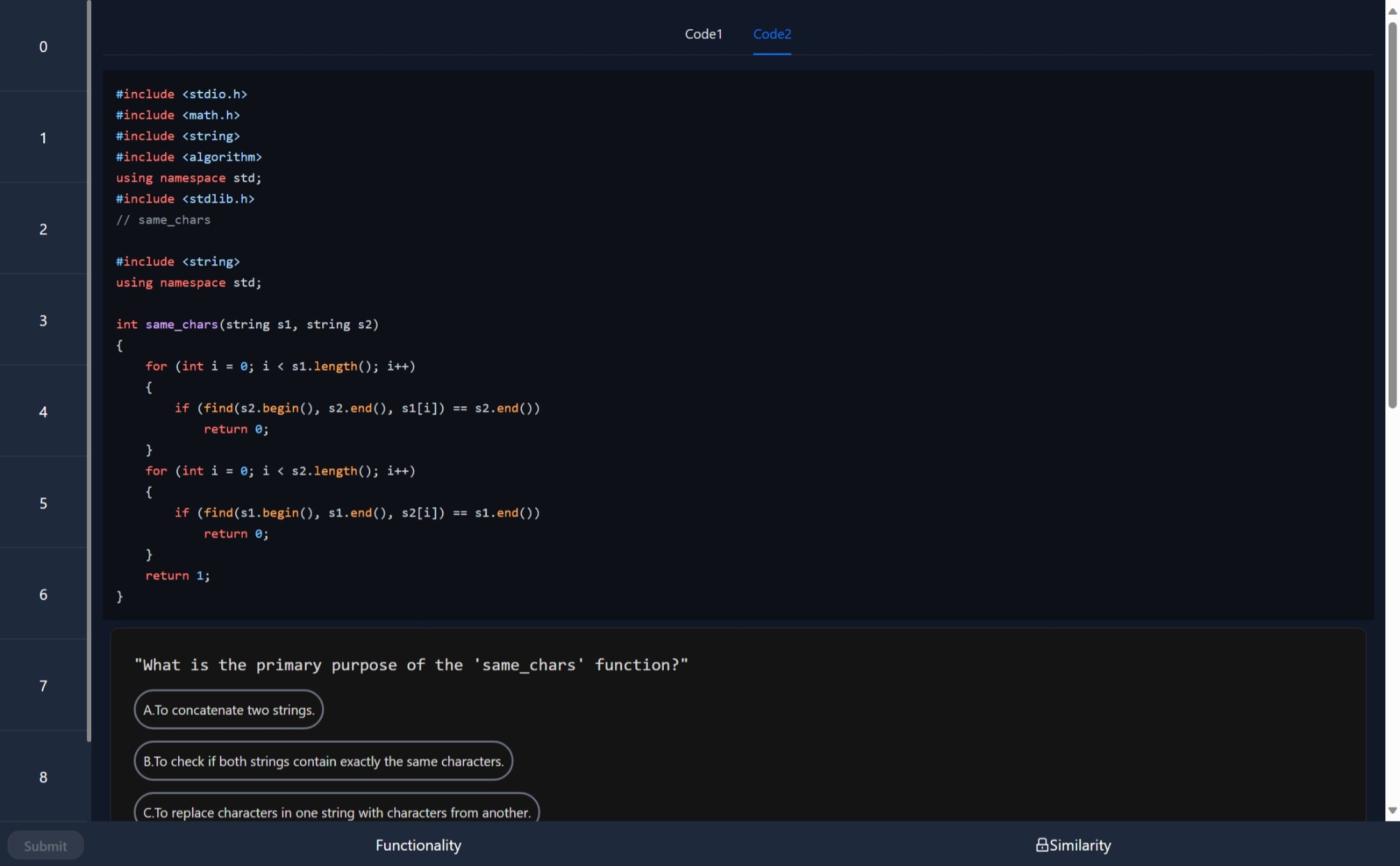}
    \caption{Evaluation page.}
    \label{fig:evaluation_page}
\end{figure}
Then, in the actual evaluation page as shown in \ref{fig:evaluation_page}, each participant is provided with 10 sets of questions. The 'Submit' button remains inactive until all sections are completed, ensuring the completion of the study.

%%%%%%%%%%%%%%%%%%%%%%%%%%%%%%%%%%%%%%%%%%%%%%%%%%%%%%%%%%%%%%%%%%%%%%%%%%%%%%%
%%%%%%%%%%%%%%%%%%%%%%%%%%%%%%%%%%%%%%%%%%%%%%%%%%%%%%%%%%%%%%%%%%%%%%%%%%%%%%%

\end{document}